\documentclass[11pt,a4paper]{article}
\usepackage{jcappub}
\usepackage{natbib}

\usepackage{multirow}
\bibpunct{(}{)}{;}{a}{}{,}
\usepackage{hyperref}
\usepackage{amsmath}
\hypersetup{colorlinks=true,citecolor=blue}
\usepackage{graphicx}
\usepackage[T1]{fontenc}
\usepackage{amsmath}
\usepackage{amssymb}
\usepackage{amsmath}
\usepackage{amsfonts}
\usepackage{epsfig,graphics}
\usepackage{epstopdf}

\newcommand{\mnras}{Monthly Notices of the Royal Astronomical Society}
\newcommand{\apjl}{Astrophysical Journal Letters}
\newcommand{\prd}{Physical Review D}
\newcommand{\apjs}{Astrophysical Journal, Supplement Series}
\newcommand{\apj}{Astrophysical Journal}
\newcommand{\aap}{Astronomy and Astrophysics}
\newcommand{\jcap}{Journal of Cosmology and Astroparticle Physics}
\newcommand{\aj}{Astronomical Journal}

\newcommand{\bea}{\begin{eqnarray}}
\newcommand{\eea}{\end{eqnarray}}

\title{Planck CMB Anomalies: Astrophysical and Cosmological Secondary Effects and the Curse of Masking} 
\author[a]{A. Rassat}
\author[b]{J.-L. Starck}
\author[b]{P. Paykari}
\author[b]{F. Sureau}
\author[b]{J. Bobin}

\affiliation[a]{Laboratoire d'Astrophysique, Ecole Polytechnique F\'ed\'erale de Lausanne (EPFL), Observatoire de Sauverny, CH-1290, Versoix, Switzerland.}
\affiliation[b]{Laboratoire AIM, UMR CEA-CNRS-Paris, Irfu, SAp, CEA Saclay, F-91191 GIF-SUR-YVETTE CEDEX, France.}

\emailAdd{anais.rassat@epfl.ch}

\abstract{Large-scale anomalies have been reported in CMB data with both WMAP and Planck data. These could be due to foreground residuals and or systematic effects, though their confirmation with Planck data suggests they are not due to a problem in the WMAP or Planck pipelines. If these anomalies are in fact primordial, then understanding their origin is fundamental to either validate the standard model of cosmology or to explore new physics. We investigate three other possible issues: 1) the trade-off between minimising systematics due to foreground contamination (with a conservative mask) and minimising systematics due to masking, 2) astrophysical secondary effects (the kinetic Doppler quadrupole and kinetic Sunyaev-Zel'dovich effect), and 3) secondary cosmological signals (the integrated Sachs-Wolfe effect).  We address the masking issue by considering new procedures that use both WMAP and Planck to produce higher quality full-sky maps using the sparsity methodology (LGMCA maps). We show the impact of masking is dominant over that of residual foregrounds, and the LGMCA full-sky maps can be used without further processing to study anomalies.  We consider four official Planck PR1 and two LGMCA CMB maps. Analysis of the observed CMB maps shows that only the low quadrupole and quadrupole-octopole alignment seem significant, but that the planar octopole, Axis of Evil, mirror parity and cold spot are not significant in nearly all maps considered. After subtraction of astrophysical and cosmological secondary effects, only the low quadrupole may still be considered anomalous, meaning the significance of only one anomaly is affected by secondary effect subtraction out of six anomalies considered. In the spirit of reproducible research all reconstructed maps and codes will be made available for download here \url{http://www.cosmostat.org/anomaliesCMB.html}.}

\begin{document}

\maketitle
\flushbottom

\section{Introduction}
Recent cosmological data \citep[e.g.][]{kilbinger2013,BOSS:DR11,PlanckCP} provide strong support for the standard model of cosmology. In this model, fluctuations in the early Universe are thought to arise from a period of accelerated expansion called inflation, during which quantum mechanical fluctuations are stretched to cosmological scales. The temperature fluctuations observed in the Cosmic Microwave Background (CMB) are related to these early fluctuations and are thus thought to obey two fundamental properties: statistical isotropy and Gaussianity.

However, several independent observations of the large-scale CMB have reported violations of statistical isotropy or Gaussianity, dubbed "anomalies". A low quadrupole was reported in COBE data \citep{Hinshaw:1996,ml:bond98} and confirmed with WMAP data \citep{Spergel:2003cb,w9:lowquad} as well as lack of correlations on large scales \citep{Spergel:2003cb,Copi:2007}. Other anomalies have been reported in WMAP data. The octopole presented an unusual planarity and its phase seemed correlated with that of the quadrupole \citep{Tegmark:2003ve,OctPlanarity,Slosar:2004s,Schwarz:2004,Copi:2005ff,Copi:2007,Copi:2010,Rassat:2012}. Other anomalies include a north/south power asymmetry \citep{Eriksen:2003db,Bernui:2006,Hansen:2009}, an anomalous cold spot \citep{Vielva:2004,wave:vielva04,gauss:cruz05,McEwenCS2005,Cruz:2006sv,Vielva2010}, the so-called `Axis of Evil' \citep[AoE,][]{Copi:aoe,Land:2005ad,Land:aoe2,Rassat:axes}, anomalous alignments of other large-scale modes \citep{Bielewicz:2004} and other violations of statistical isotropy \citep{Hajian:2003s,Land:2004bs,Kim:2010,Kim:2010b,Kim:2011,Rassat:axes}.  Other measures of non-Gaussianity were also reported \citep{Eriksen:2004,Eriksen:2005,Rath:2007,Rath:2009}.

Anomalies have also been reported in \emph{Planck} data \citep{PlanckStat,Copi2013powelack}, mainly regarding lack of statistical isotropy (quadrupole-octopole alignment, low variance, hemispherical asymmetry, phase correlations, power asymmetry, dipolar modulation, generalised power modulation, parity asymmetry), with little evidence of non-Gaussianity except for the cold spot \citep{coldspot:planck}.

Understanding the statistical significance of these anomalies and their potential source(s) is of utmost importance for cosmology, since lack of statistical isotropy or Gaussianity on large-scales in the primordial CMB would be problematic for the validation of our standard model of cosmology. There are several possible causes to these anomalies. The most exciting would be exotic early Universe physics \citep[e.g.,][]{anomalies:exotic,anomalies:exotic2,anomalies:exotic3,anomalies:exotic4,exotic:sn}, though other explanations are also possible: the anomalies could be a simple statistical fluke, i.e. we happen to live in a realisation where chance alignments/anomalous signatures exist on large scales. Other studies have concluded that posterior statistics increase the significance of these anomalies, and that they are in fact not anomalous, as concluded by the WMAP team with their ninth data release \citep[][]{WMAP9map}. There could be some problem with the CMB data which could lead to observed large-scale anomalies, though the fact that these have been observed in the two independent experiments of WMAP and Planck seems to rule out a potential problem with the data.

The goal of this paper is to investigate three other possible issues which could be related to the reported anomalies:
\begin{enumerate}
\item Confusion and obscuration due to our Galaxy, and other Galactic foregrounds mean the largest scales usually require some level of mask processing. This could bias large-scale statistics either because of foreground residuals or because of the mask processing itself.

\item Astrophysical secondary effects could be linked to the anomalies, especially those related to our Galaxy or Halo: e.g., the kinetic Sunyaev-Zel'dovich effect \citep{Rubin:ksz} or to our motion with respect to the CMB rest frame \citep[kinetic Doppler effect,][]{Schwarz:2004,Copi:2005ff}, since these local effects are projected to large scales on the CMB sky.

\item The removal of known secondary large-scale cosmological signals was shown to reduce the significance of some anomalies \citep*[e.g., the Integrated Sachs-Wolfe effect,][]{Francis:2010iswanomalies,Rassat:2012,Rassat:axes}. We note that this does not necessary mean the secondary signals are causing the anomalies (see Section \ref{sec:secondary} for a discussion of this). 

\end{enumerate}

\section{Large-scale anomalies studied in this paper}\label{sec:anomalies}
In this paper we consider the same reported large-scale anomalies as in \cite*{Rassat:2012} and \cite{Rassat:axes} with the addition of the cold spot, i.e.: \begin{itemize}
\item the low quadrupole, 
\item the quadrupole/octopole alignment, 
\item the planarity of the octopole, 
\item the Axis of Evil, 
\item mirror parity, 
\item the cold spot.
\end{itemize} References of statistical details of how to measure and assess the significance of each of these anomalies is given in Appendix \ref{app:anomalies}. As discussed in the previous section, the statistical significance of some of these anomalies is debated or has decreased with more recent data, e.g. the quadrupole/octopole alignment (see \cite{WMAP9map} which finds it is not anomalous compared with \cite{Copi2013alignements} who find it anomalous), the low quadrupole \citep[][]{WMAP9map} or the planar octopole \citep[which seems to be no longer anomalous since WMAP third year data, see][]{Rassat:2012}. We still consider them as part of our analysis to avoid posterior statistics, i.e. focussing on statistics that return anomalous behaviours.

\section{Mask processing and choice of mask} \label{sec:processing}

Confusion, obscuration and other foreground emissions due to our Galaxy mean that CMB observations along the Galactic plane and bulge may be incomplete or contain residuals. In addition, Galactic foregrounds can contaminate the CMB temperature observations. Masking the corresponding pixels and processing them has traditionally been used in CMB surveys for cosmological analyses and the study of large-scale anomalies. While some studies have concluded that the claimed anomalies were stable with respect to component separation algorithms and mask choice \citep[e.g.][]{Copi2013alignements}, others have concluded that mask processing was the limiting factor of large-scale anomaly studies \citep{HarmonicInpainting,Copi:bias,WmapAnomalies,WMAP9map,Starck:2013,ParityPR12014,Planck:NSNG} which is why we investigate mask processing and choice of mask further in this paper. 

\subsection{Mask processing in the literature}

There exists a wide variety of mask processing methods in the literature today, and we briefly review here the different approaches.

The simplest approach is either to ignore masked pixels \citep[e.g.,][]{Land:2005ad,OctPlanarity,Rassat:2007krl} or to set their values to zero, though this may create an artificial signal and leakage between modes which then must be included in the covariance matrix. Others have proposed diffuse inpainting, where the masked pixels are replaced with average values (as done for two of the official Planck maps \citep{PR1_compsep} or in \cite{Francis:2010iswanomalies} with CMB and 2MASS data). In this case, statistics in the mask may differ from statistics in the true underlying CMB. Assuming the underlying CMB is an isotropic Gaussian random field, one can use constrained Gaussian realisations \citep[e.g.,][]{bucher2012,kim2012,PavelInpainting}. This method was also used in the Planck PR1 press release \citep{pr1PressReleasse}. However,  this method may destroy existing anomalies due to isotropy and Gaussianity assumptions. A Wiener filtering method has been used \citep{OliveiraTegmark06,peiris2011,david2012,PlanckStat}, though this method assumes isotropy and an input cosmology, and \cite{Copi2013alignements} showed that it could affect anomalies. 

Recently, sparse inpainting methods  have been applied to cosmological and astrophysical data, and have been shown to be useful for weak lensing analyses \citep{starck:pires08,pires10}, Fermi data \citep{starck:schmitt2010} and asteroseismic data \citep{sato2010}. These have also been used to study the CMB \citep{inpainting:abrial06,starck:abrial08} and more specifically the large-scale CMB modes  and anomalies \citep{Dupe:2010,Rassat:2012,Rassat:axes,Starck:2013,starck2013Bayes}. Sparsity-based inpainting was shown not to destroy CMB  weak-lensing, ISW signals or some large-scale anomalies in the CMB \citep{perotto10,Dupe:2010,Rassat:2012,Rassat:axes,Starck:2013}. 

An isotropy prior was proposed in \cite{Starck:2013}, but was shown not to be as efficient as existing sparse inpainting methods. An energy prior was also used \citep{Starck:2013}, however this method assumes both isotropy and an input power spectrum.

We stress that there exists a wide range of mask processing methods with different priors and even algorithms, and that understanding the prior alone is not enough to predict the quality of a given reconstruction method. In the current state of the art, we think it is wise to consider that several inpainting methods may be useful, each having pros and cons in the context of large-scale CMB analysis. In this paper we focus on studying sparse inpainting reconstruction methods for large-scale CMB analysis.

\subsection{Do we still need masking ?}

CMB experiments have often provided several masks with different fractional sky coverage values (`$f_{\rm sky}$'), and in addition to the method used for mask processing, the choice of mask itself can affect the final analysis. The fundamental idea is that there is a trade-off between minimising systematics due to foreground contamination (by choosing a conservative or `aggressive' mask) and minimising systematics due to loss or bias of large-scale information and features (by choosing a more optimistic mask). 

Recent studies on the quadrupole-octopole alignment have performed the analysis with different masks and processing choices. The Planck team used the U73 mask \citep[$f_{\rm sky}=0.73$,][]{PR1_compsep} with a Wiener filter, while \cite{Copi2013alignements} used an optimistic mask ($f_{\rm sky}=0.97$) with a harmonic inpainting method, and both of these papers concluded on the presence of the quadrupole-octopole alignment. \cite*{Rassat:2012} found that using sparse inpainting with the WMAP7 temperature analysis mask reduced the significance of the quadrupole-octopole alignment so that it was no longer significant.  For other statistics the choice of mask and processing was shown to introduce biases \cite[see for e.g.][for mirror parity studies]{ParityPR12014}. In \cite{PlanckStat}, a curious behaviour of CMB map wavelet coefficients was found: 
the wavelet coefficients distribution was more anomalous using the CG60 mask than with the CG70 mask, contrarily to what was expected.

The question is whether it is possible with current data to do without masking altogether?  \cite{Copi:bias} concluded that despite the information loss, using the unobscured region without reconstructing data within the Galactic mask may be a more robust measure of the true CMB, though they also concluded that in realistic cases with noise and residuals, treating the data in this way may lead to \emph{highly biased reconstructions}'.

However, recent Planck data has a better resolution and more channels than WMAP, and can be combined with WMAP data ensuring better foreground removal than using Planck or WMAP data alone. New CMB maps derived using Planck PR1 and  WMAP-9yr data jointly are even full-sky without any visible foreground residual in the Galactic plane \citep[LGMCA,][]{BobinWPR1_2014}.  \cite{ParityPR12014} applied mirror parity analysis to the LGMCA map and concluded that the full-sky map could be used without further mask processing.

In the following subsection, we investigate quantitatively with simulations whether masking is still necessary with the latest Planck and WMAP data and the latest component separation methods. 

\subsection{Testing the trade-off between optimistic and aggressive masking}\label{sec:mask:test}
\begin{figure*}[htb]
\vbox{
\hbox{
\includegraphics[scale=0.27]{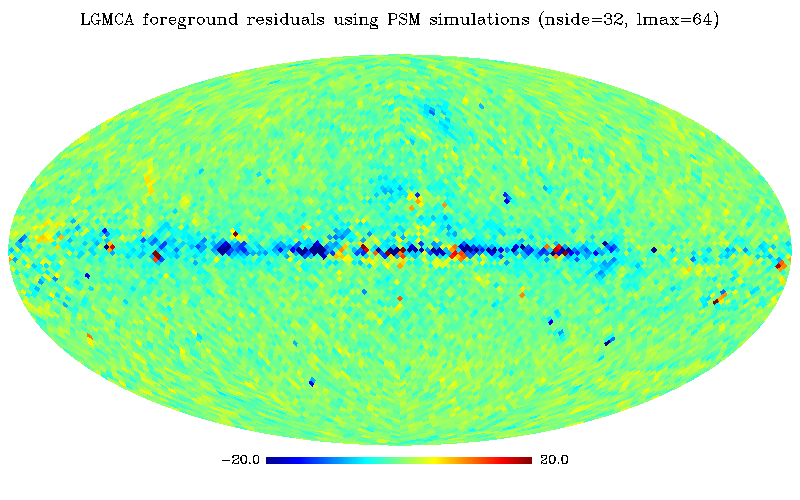}
\includegraphics[scale=0.27]{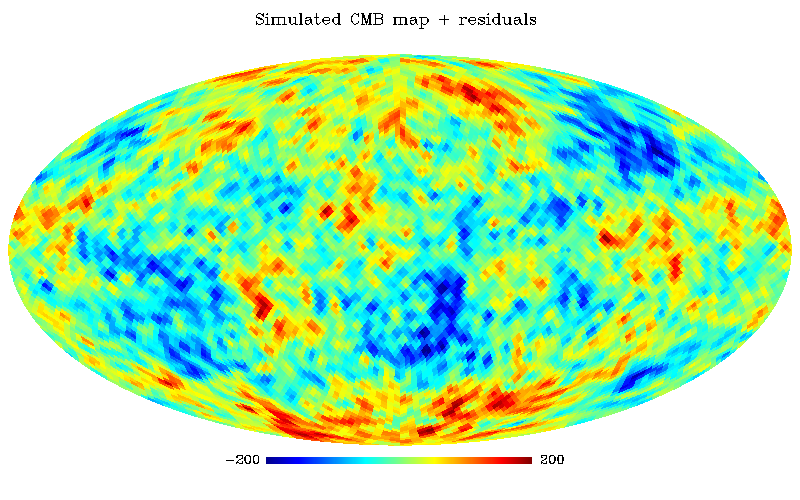}
}}
\caption{\emph{Left}: Estimated error map  $\epsilon$ (see Equation \ref{eq:errormap}) at nside=32 on PSM data using LGMCA component separation.  \emph{Right}: a CMB realisation temperature map with added foreground residuals (\emph{right}). All maps are in $\mu K$.}
\label{fig_simucmb}
\end{figure*}

We first estimate a realistic foreground residual map for the CMB sky. To do this, we simulate a WMAP and Planck data set ($\delta^{T,\rm PSM}$) using the Planck Sky Model \citep[PSM,][see \ref{sec:simupsm} for details of the simulation]{PSM12}. We then apply the LGMCA component separation method (i.e., a joint WMAP-Planck analysis) with the Planck best fit ($\Omega_b h^2=0.022068, \Omega_{\rm cdm}h^2=0.12029, \theta=1.04122, \tau=0.0925, A_s=2.215.10^{-9}, n_s= 0.9624$) and obtain a foreground cleaned map ($\delta^{T, \rm LGMCA}$)\footnote{Codes and scripts are available  here: \url{http://www.cosmostat.org/planck_wpr1.html}}. We then calculate the error map $\epsilon$, given by: 
\begin{equation} \epsilon = \delta^{T,\rm PSM}-\delta^{T, \rm LGMCA},\label{eq:errormap}\end{equation}
which contains both foreground residuals and realistic noise. The estimated error map is shown in Figure \ref{fig_simucmb} (\emph{left}) with nside=32, though it was calculated using the full resolution (nside=2048).

\begin{figure*}[htb]
\vbox{
\hbox{
\includegraphics[scale=0.27]{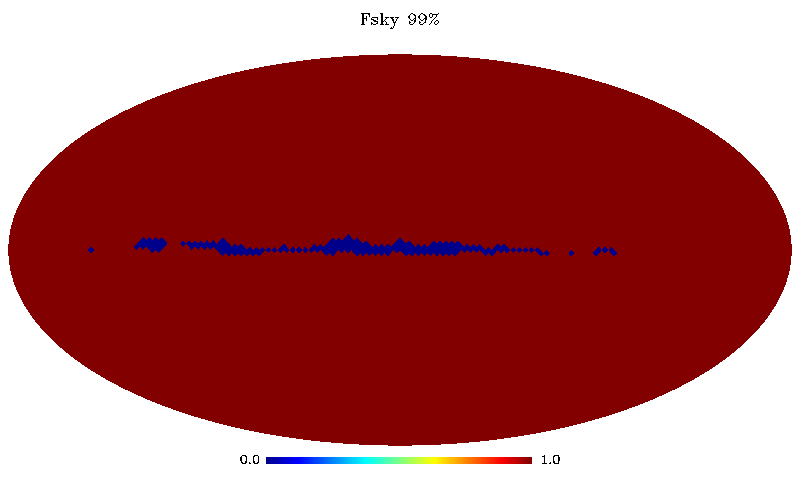}
\includegraphics[scale=0.27]{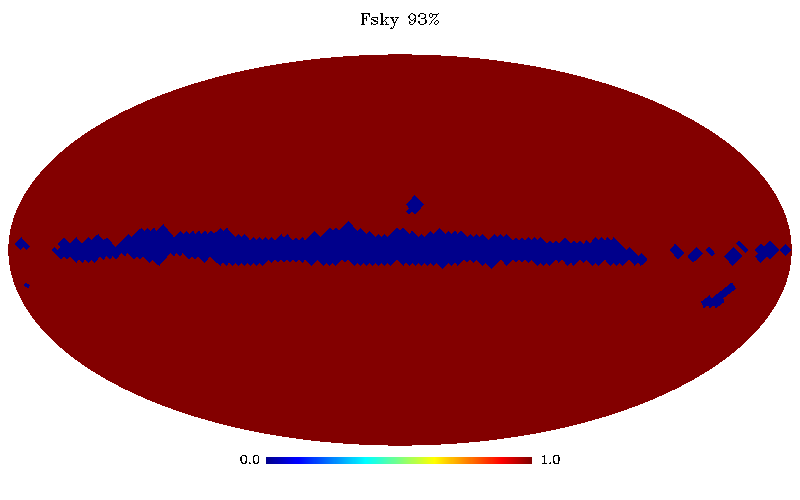}
}
\hbox{
\includegraphics[scale=0.27]{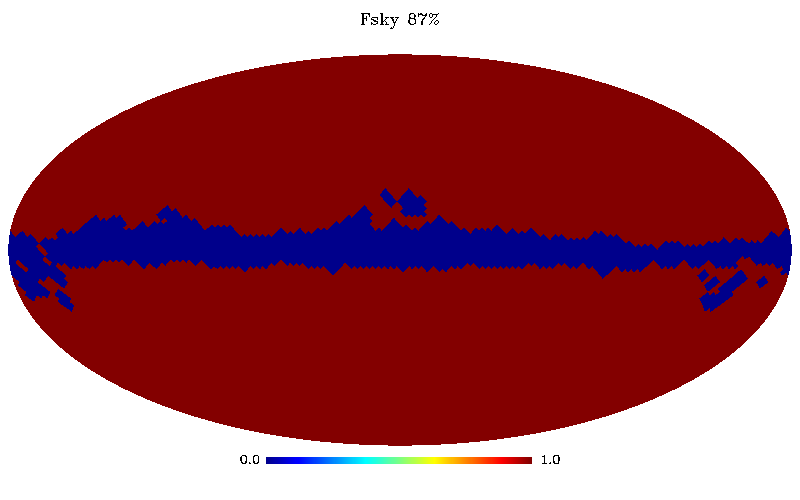}
\includegraphics[scale=0.27]{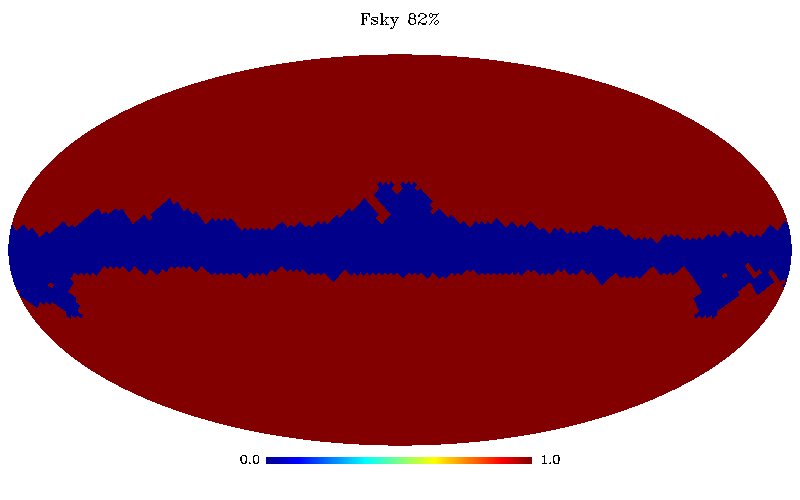}
}
\hbox{
\includegraphics[scale=0.27]{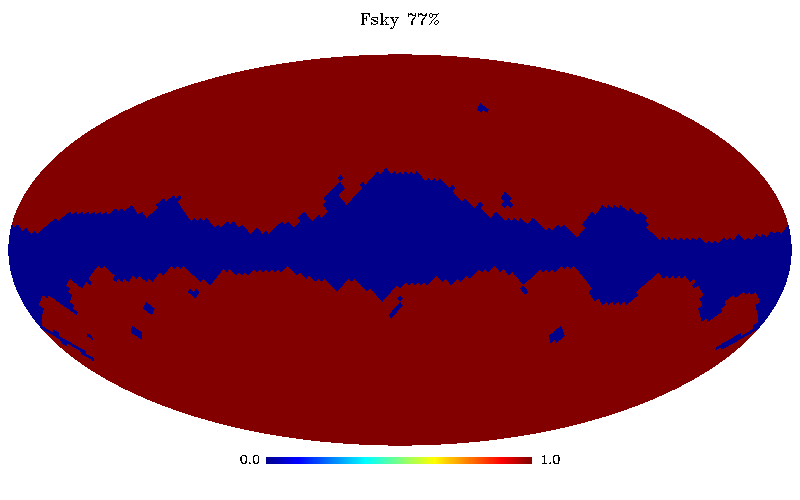}
\includegraphics[scale=0.27]{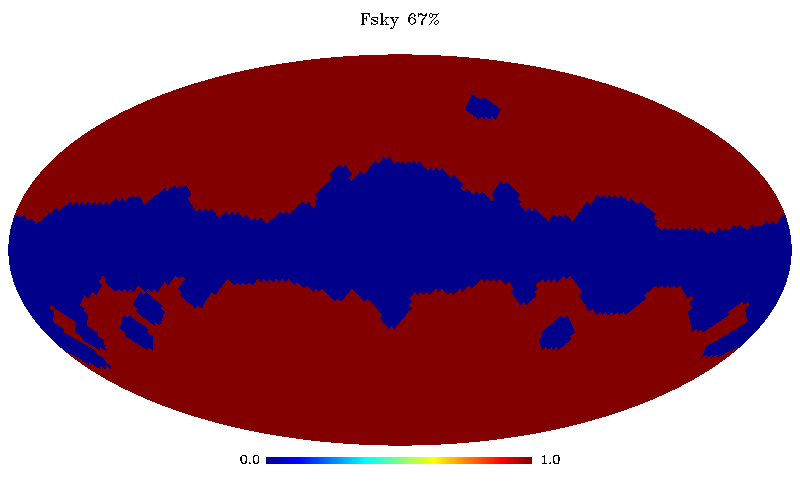}
}
\hbox{
\includegraphics[scale=0.27]{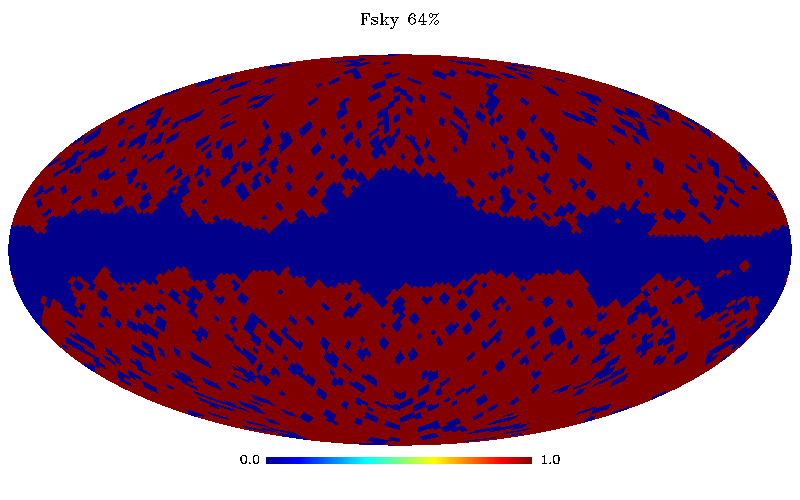}
\includegraphics[scale=0.27]{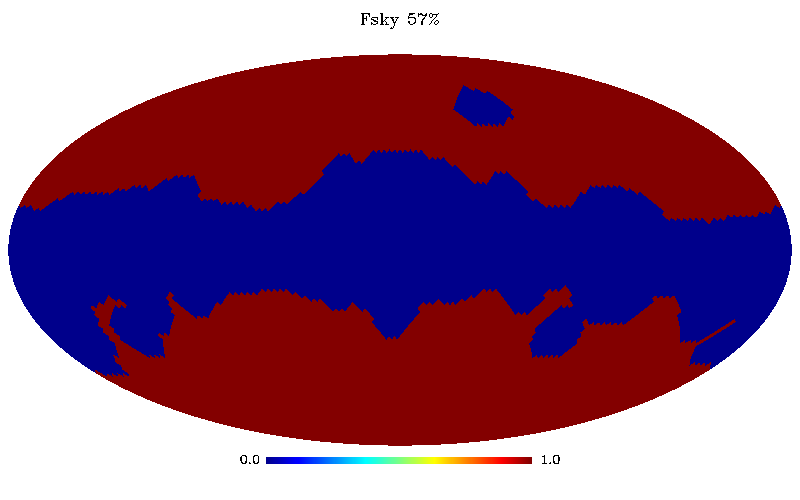}
}}
\caption{Masks tested in this paper with respective values of $f_{\rm sky}$ (from top to bottom, left to right): 0.99, 0.93, 0.87, 0.82, 0.77, 0.67, 0.64 and 0.57.}
\label{fig_mask}
\end{figure*}

To study the trade-off between optimistic vs. aggressive masking, we perform the following test: 
\begin{itemize}
\item Generate $2\times n$ CMB realisations, one set with only temperature data (i.e., a simulated `true CMB'), and another with the residual $\epsilon$ map added (i.e., a simulated `observed' CMB after LGMCA component separation).  An example CMB realisation with the error map added is shown in Figure \ref{fig_simucmb} (\emph{right}), showing the error map is low in amplitude compared to the temperature  signal.
\item For each set of CMB realisations, we apply sparse inpainting for 9 levels of masking, corresponding to $f_{\rm sky}=\{0.57, 0.64, 0.67, 0.77, 0.82, 0.87, 0.93, 0.99, 1\}$. For the last value $f_{\rm sky}=1$ we use the full-sky map as is and do not apply inpainting. Figure~\ref{fig_mask} shows the used masks. 
Note the 64\% mask contains many point sources and a slightly thinner galactic plane than
the 67\% mask, in order to also test the effect of point source masking. 

\end{itemize} For $n=200$ we therefore consider a total of 3600 maps, i.e. 400 full sky maps (with and without residuals) and 3200 maps with different levels of masking. 

\begin{figure*}[htb]
\vbox{
\hbox{
\includegraphics[scale=0.18]{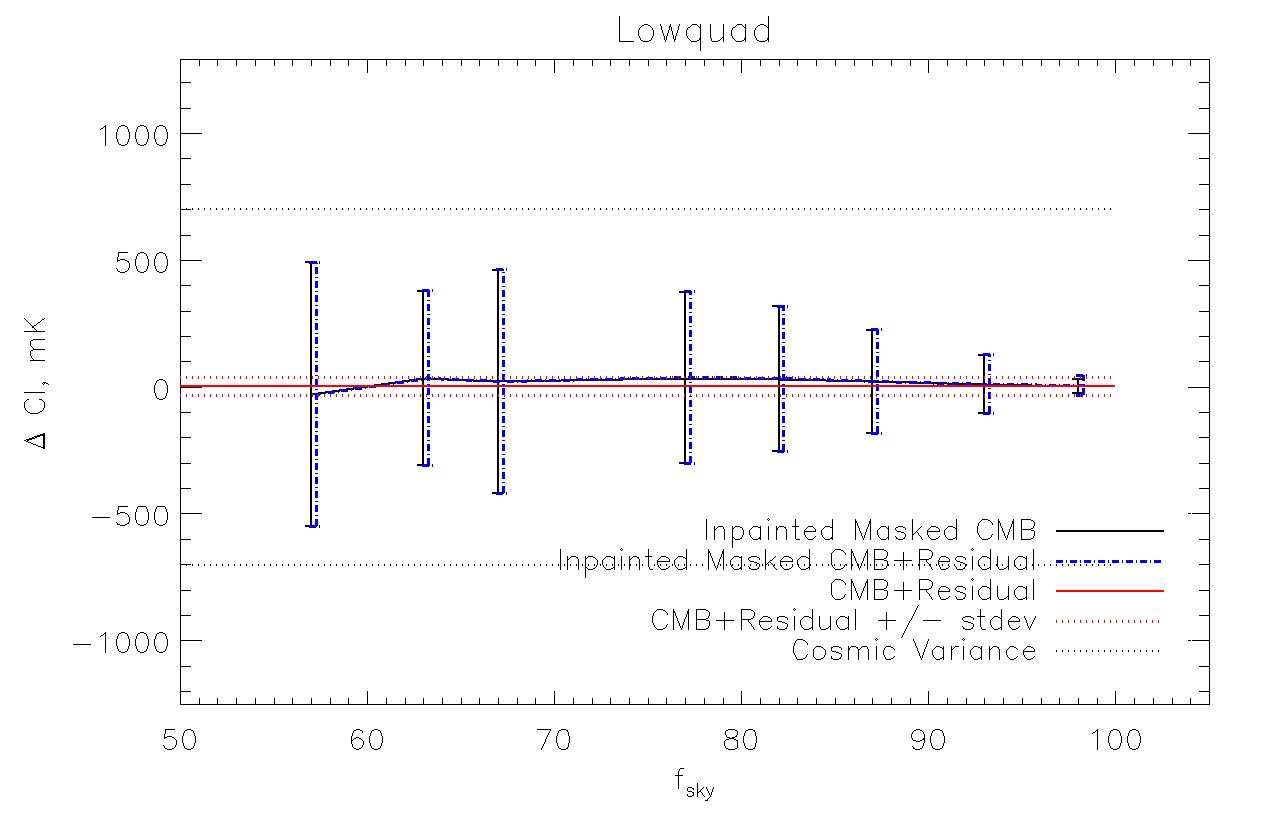}
\includegraphics[scale=0.18]{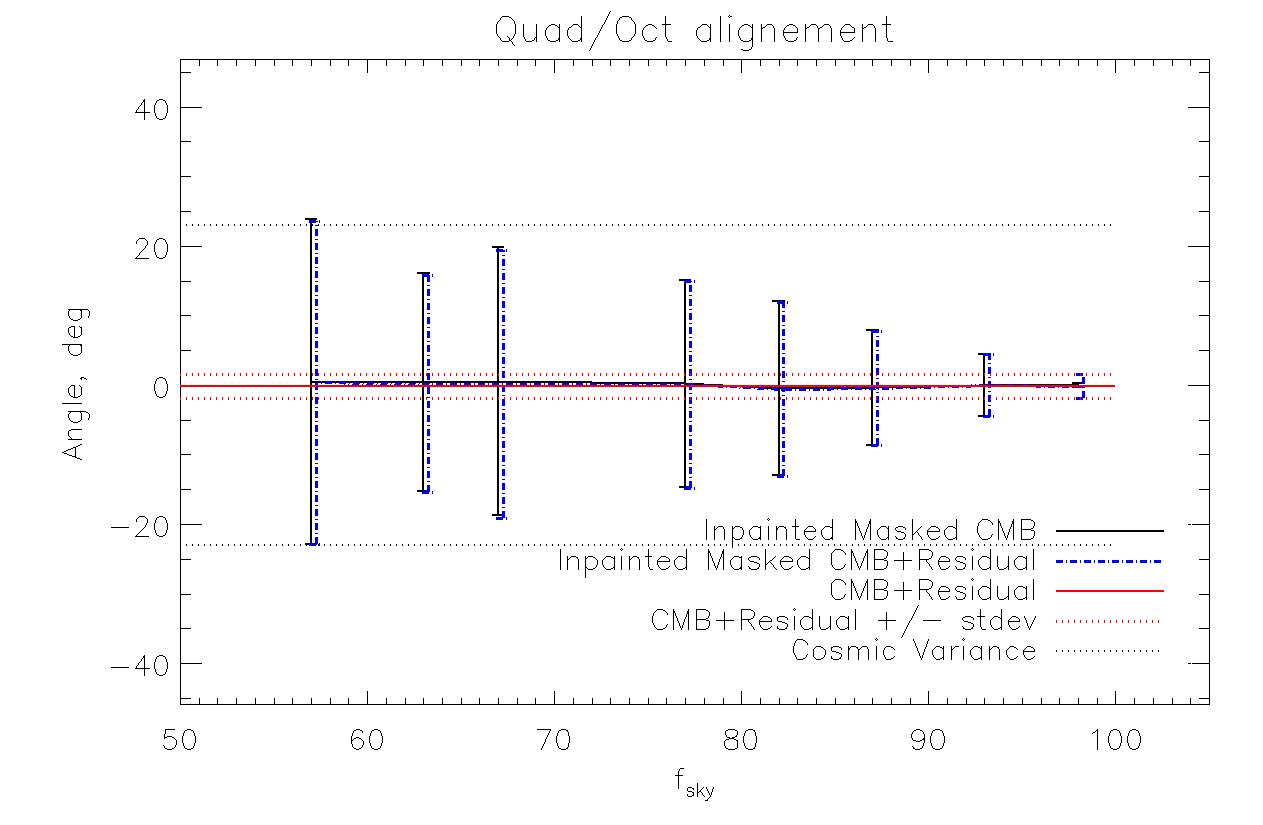}
}
\hbox{
\includegraphics[scale=0.18]{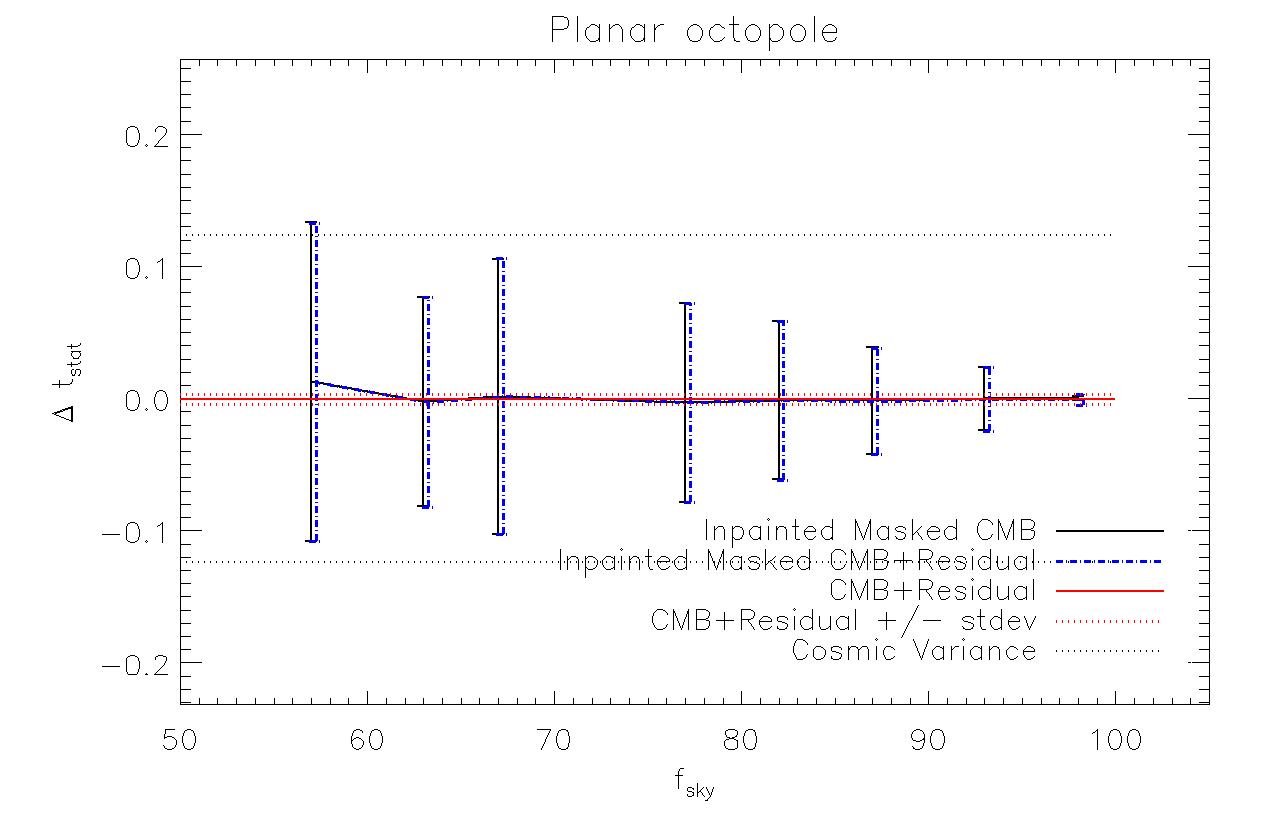}
\includegraphics[scale=0.18]{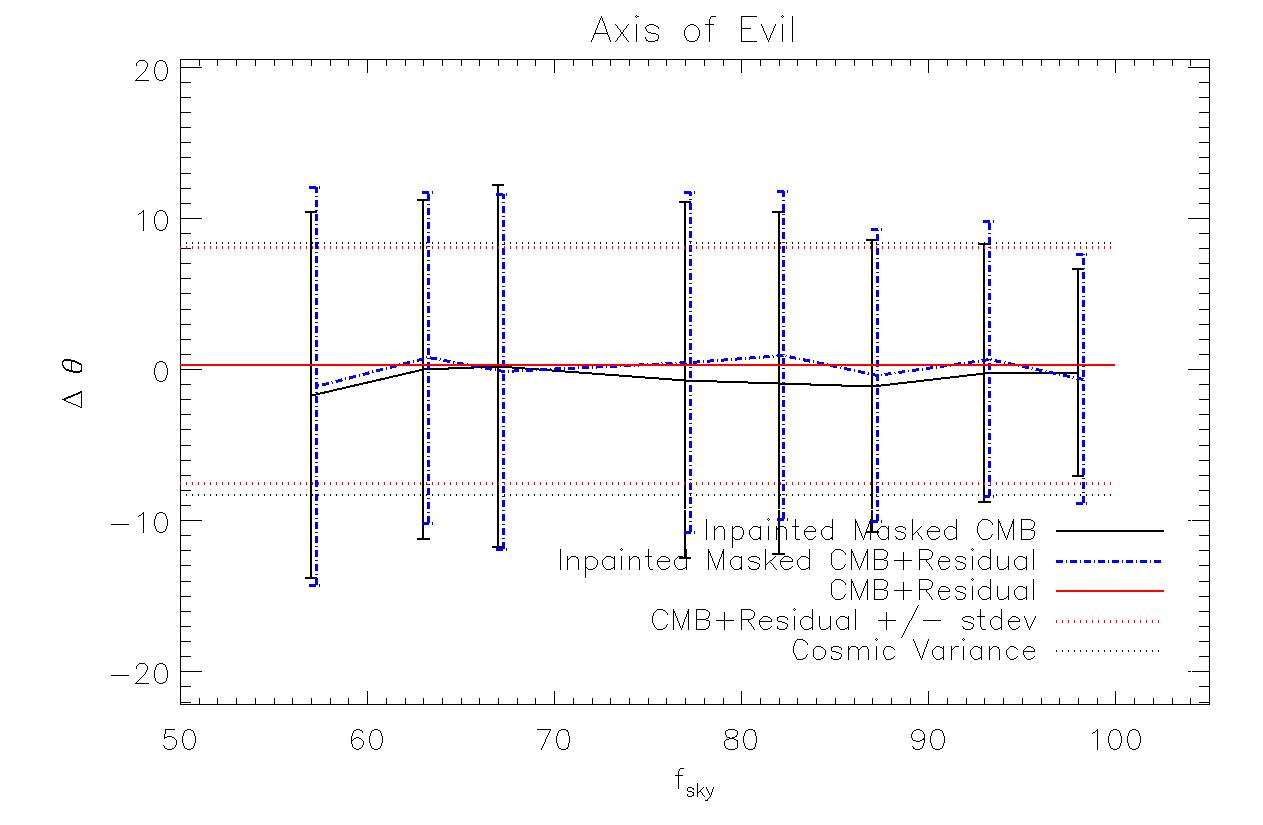}
}
\hbox{
\includegraphics[scale=0.18]{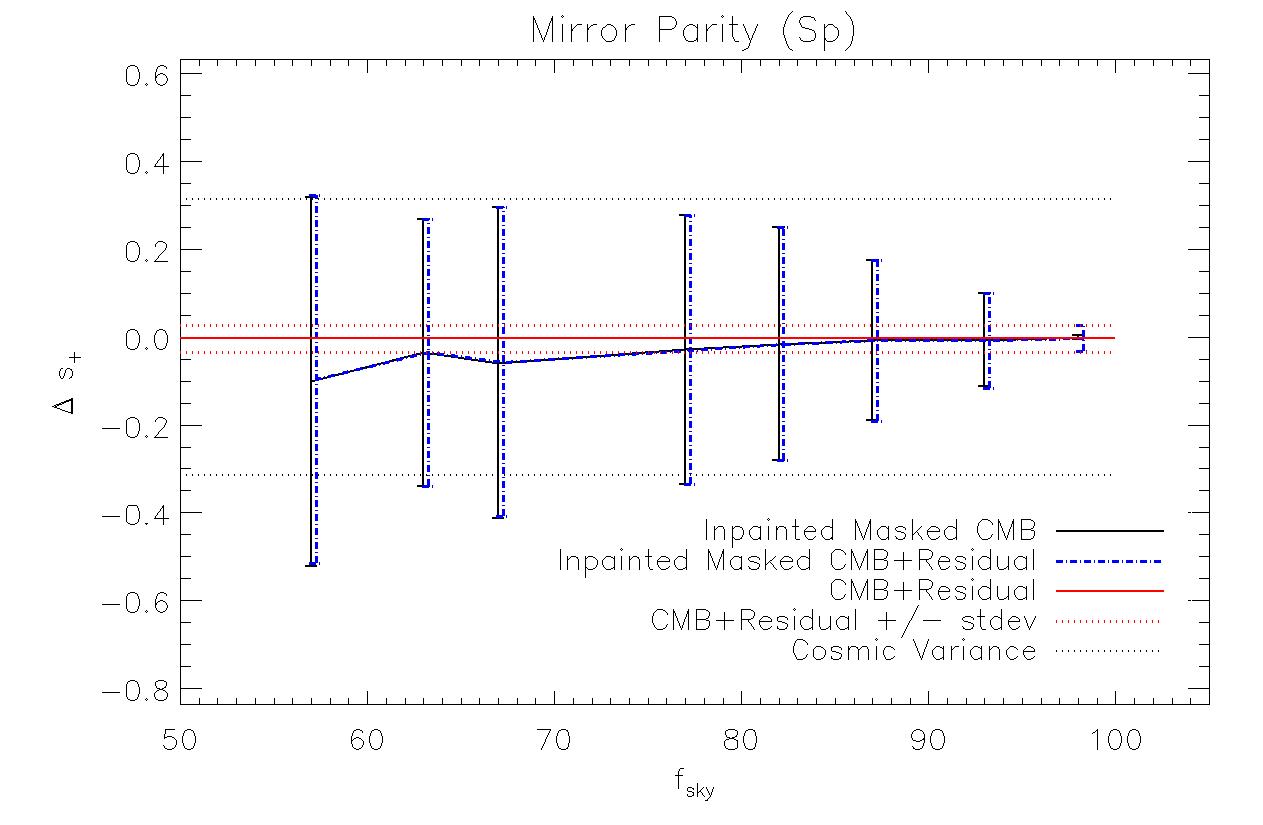}
\includegraphics[scale=0.18]{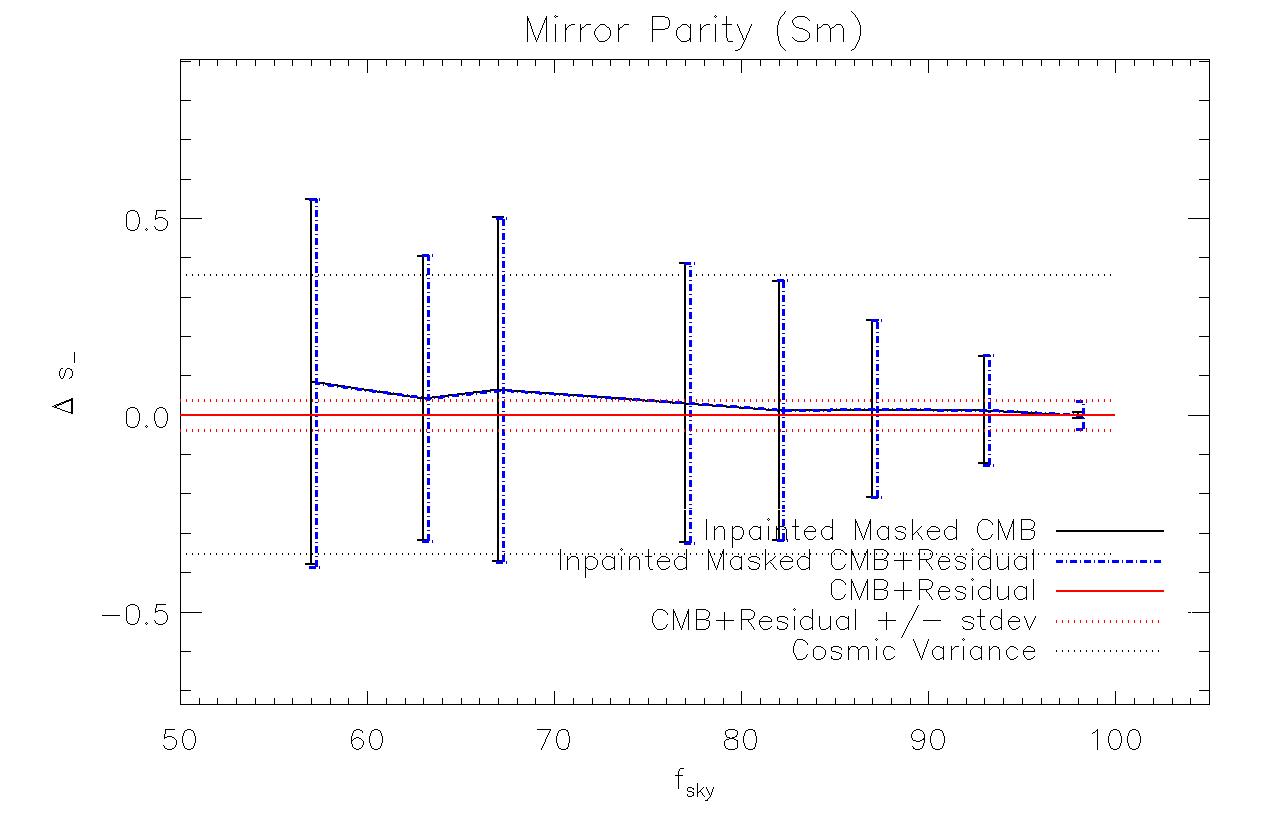}
}}
\caption{Mean difference ($\Delta p_i$) different statistics $\{p_i\}$ as measured for the full-sky CMB maps versus those measured after inpainting, as a function of mask size. The statistics $\{p_i\}$ correspond to (\emph{clockwise from top left}): the low quadrupole, the quadrupole/octopole alignment, the Axis of Evil, negative mirror parity, positive mirror parity and the planar octopole. The solid black and dot-dashed blue lines (the latter are slightly offset for visual clarity), correspond respectively to the inpainted temperature map and the inpainted temperature with the error map included. The horizontal red lines correspond to the mean (solid red) and standard deviation (dotted red) due to the residuals alone (i.e. where no masking/inpainting has been applied), and the horizontal black dotted lines correspond to the cosmic variance of each statistic for a Gaussian random field}
\label{fig_inp_versus_residual}
\end{figure*}

In Figure~\ref{fig_inp_versus_residual}, we plot the mean difference and standard deviation of different statistics as measured for the full-sky CMB maps versus those measured after inpainting, as a function of mask size. The solid black and dot-dashed blue lines (the latter are slightly offset for visual clarity), correspond respectively to the inpainted temperature map and the inpainted temperature with the error map included. The horizontal red lines correspond to the mean (solid red) and standard deviation (dotted red) due to the residuals alone (i.e. where no masking/inpainting has been applied), and the horizontal black dotted lines correspond to the cosmic variance of each statistic for a Gaussian random field.

From Figure \ref{fig_inp_versus_residual}, we conclude systematics due to residuals from the LGMCA component separation method are negligible compared to cosmic variance and compared to the impact of masking.  The conclusion is that - at least for the LGMCA method - one should prefer the most optimistic mask (i.e., no mask) to an aggressive one since the resulting contamination by foregrounds is not problematic for the study of large-scale anomalies in the CMB. In this case, it is the conservative approach of masking incomplete or contaminated data which does a disfavour to the final statistics of the reconstructed CMB map.  In this paper, we therefore consider LGMCA CMB maps without any mask, and the official Planck PR1 maps with their respective masks.

\subsection{Wavelet analysis and masking for the cold spot}\label{sec:coldsot:mask}

We have excluded the cold spot anomaly from the tests presented in Figure \ref{fig_inp_versus_residual}, since we are interested in investigating it with wavelets and not spherical harmonics. Considering large wavelet scales obtained from a masked image requires that the wavelet coefficients be analysed with a second mask, much more aggressive than the first one. This is due to the fact that wavelet coefficients close to the mask edge are impacted by the masking. The larger the wavelet scale is, the more aggressive the second masking needs to be. If one is interested in the cold spot, a second mask corresponding to $20\%$ of the full-sky area has to be removed from the analysis, so that the final $f_{\rm sky}$ value is $\sim 0.5$.

\begin{figure*}[htb]
\vbox{
\hbox{
\includegraphics[scale=0.24]{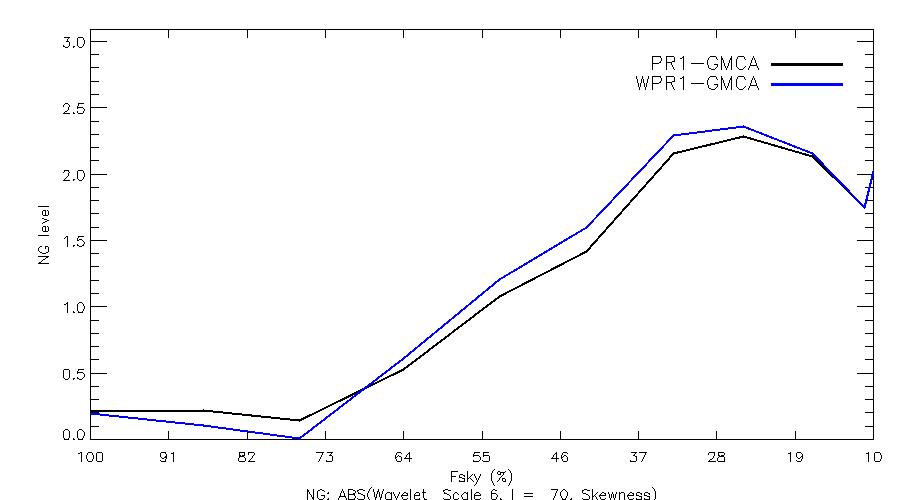}
\includegraphics[scale=0.24]{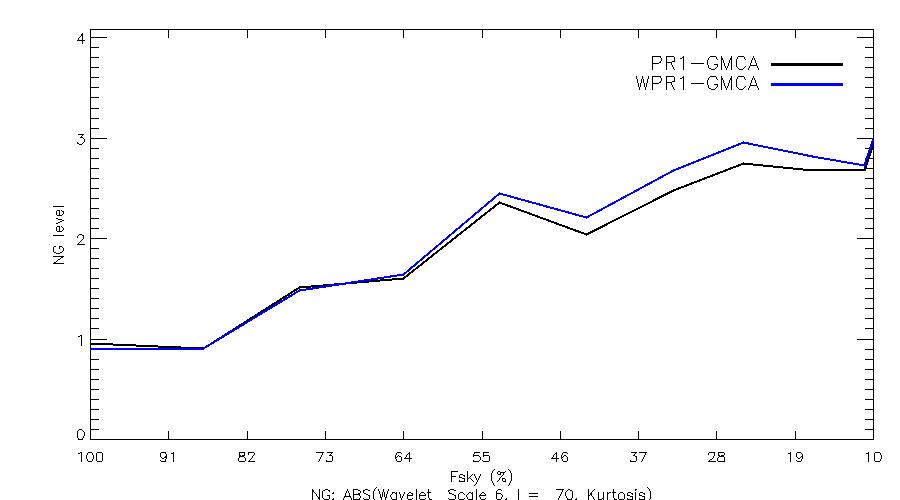}
}
\hbox{
\includegraphics[scale=0.24]{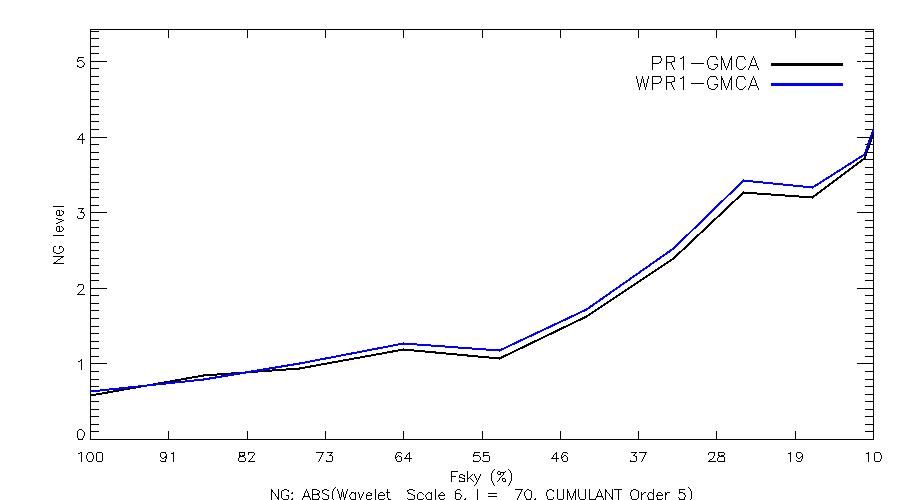}
\includegraphics[scale=0.24]{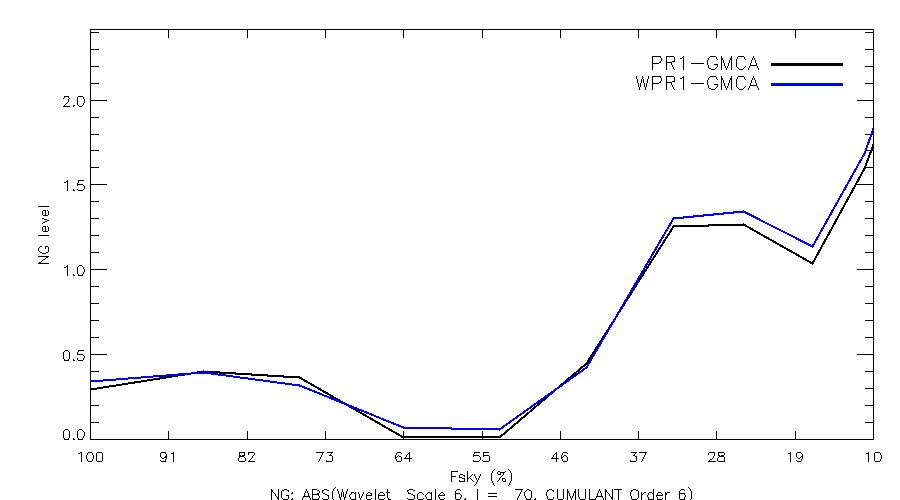}
}}
\caption{High order statistics of a large scale wavelet band, centred around $l=70$, vs. sky coverage, for two full-sky CMB maps, PR1-LGMCA (black line) and WPR1-LGMCA (blue line).  A given $f_{\rm sky}$ on the $x$-axis corresponds to a surface area of the sky with a Galactic latitude $|b|<L_i$ as described in \ref{sec:coldsot:mask}. \emph{Top}: skewness and kurtosis vs. sky coverage; \emph{bottom}: cumulants of order 5 and 6 vs. sky coverage. The values are normalised by those obtained from noise realisations so that the y-axis can be seen as a detection level. }
\label{fig_coldspot_hos_versus_fsky}
\end{figure*}

To investigate the impact of masking on CMB wavelet coefficient statistics, we apply a wavelet transform to 
the two full-sky CMB maps, LGMCA-PR1 and LGMCA-WPR1.  We then calculate  high-order statistics (skewness, kurtosis,
cumulant of order 5 and 6) on the wavelet scale corresponding to the cold spot angular size,  considering 13 different mask sizes, masking parts of the sky  where the Galactic latitude $b$ ($^\circ$)  is such that $ \mid b \mid < L_i$,  $ L_i =$\{86.5, 79.6, 72.7,  65.8,  58.9,   51.9, 45.0, 38.0,  31.1, 24.2, 17.3, 10.3, 3.5\}. A set of 80 realisations of CMB maps (assuming the fiducial cosmological model obtained from the Planck PR1 results) and noise maps were also processed through the LGMCA pipeline. 
Statistics obtained from the two CMB maps were normalised by the standard deviation of the 
statistics computed from the noise realisations.
Figure \ref{fig_coldspot_hos_versus_fsky} shows these high order statistics as a function of $f_{\rm sky}$.
The normalisation by the standard deviation means the y-axis corresponds to a detection level.
We see a clear trend that each statistic becomes less anomalous as the $f_{\rm sky}$ increases, even though we would expect the opposite if the (non-Gaussian) residuals were problematic. 

Our analysis shows that localizing the statistical analysis around the location of the cold spot will automatically make it appear anomalous, thus leading us to overestimate its significance, while a full-sky analysis suggests that it is not in fact anomalous.

\section{Astrophysical and cosmological secondary effects}\label{sec:secondary}

\subsection{Observed CMB vs. primordial CMB}
The premise of this paper is that the most interesting cause of the anomalies would be one resulting from early Universe physics, and that we are therefore interested in studying the primordial CMB instead of the observed CMB, i.e., one free from Galactic emissions, secondary astrophysical and cosmological effects. 

There is a debate in the literature about several potential issues regarding this. The counter-arguments are the following \citep[see e.g.,][]{Copi2013powelack,Copi2013alignements}: 
\begin{enumerate}
\item Even if Galactic foregrounds can be removed in a satisfactory manner, other astrophysical and cosmological secondary effects are difficult to reconstruct and subject to biases;
\item If the removal of secondary cosmological effects causes the statistical significance of the anomalies to decrease, one must then explain why local effects are aligned with the primordial temperature field in such a way as to reduce the anomalies' significance. This shifts the theoretical importance of the anomalies from the primordial Universe to the late Universe. 
\end{enumerate}

Regarding the first point: in \cite*{Rassat:2012}, \cite{Rassat:axes} and in Section \ref{sec:mask:test}, we tested reconstruction of the CMB, galaxy field and ISW (by checking the power spectrum, phases and the specific statistical isotropy tests considered in this paper). We found that the reconstruction error related to sparse inpainting decreased rapidly with increasing $f_{\rm sky}$. We also found that residual Galactic foregrounds from the LGMCA method were not problematic for the statistical tests considered in this paper (see Section \ref{sec:mask:test}). With future Galactic and LSS data, reconstruction methods are expected to continue to improve, so reconstruction of secondary effects should still be attempted and tested.

Another issue is how well galaxy surveys trace the underlying gravitational potential. Possible issues are related to the galaxy bias (especially if it is scale-dependent) and completeness problems. The advantage of the reconstruction method used in this paper \citep[see Section \ref{sec:ISW} and][]{Rassat:2012,Rassat:axes} is that it is independent of the linear scale-independent galaxy bias. On the scales considered here ($\ell=2-5$), the assumption of a linear scale-independent galaxy bias is not controversial. Regarding completeness, the Galactic mask we use for 2MASS to ensure 98\% completeness in the regions considered \citep{Afshordi:2003xu,Rassat:2007krl}.

Regarding the second point, we find several arguments refuting this as a problem. The first, na\"ive, explanation is that there could be a chance alignment of primordial and secondary CMB modes. Given that we are considering very few modes, the statistical occurence of such an alignment is not small enough for its occurrence to be problematic. We note that this argument might not hold for other anomalies not considered in this paper which examine a larger range of multipoles \cite[e.g. the lack of power on large scales as studied in][]{Copi2013alignements}.

Furthermore, if the anomalies' significance is reduced after subtraction of secondary signals, this does not necessarily mean their modes are aligned with modes of the primordial CMB. The anomalies we are considering are measured with complex statistics which in some cases consider correlations between different multipoles. If we consider for example the quadrupole/octopole alignment, a change in a single multipole can be enough to break the alignment \citep[see for e.g.,][]{Rassat:2012}. Another example of this is for the reported AoE from WMAP 3rd year data, whose significance changes drastically whether or not the kinetic Doppler quadrupole ($\ell=2$) is subtracted \citep{Rassat:axes}, even though the Axis of Evil measures the scales covering $\ell=2-5$. Furthermore, in the search for preferred axes, the preferred direction for a given $\ell$ can also change if a single mode $m$ changes. If one $m$ mode in a single multipole $\ell$ is affected by subtraction of a secondary effect this can reduce the significance of an anomaly.

\subsection{Removing different types of foregrounds}
In practice, there are three types of foregrounds that can either be removed through component separation methods or by reconstructing the foreground fields and subtracting them. The three types of foregrounds are: Galactic foregrounds (i.e. emissions from our Galaxy), secondary astrophysical effects (e.g. kinetic Doppler quadrupole, thermal Sunyaev-Zel'dovich effect, kinetic Sunyaev-Zel'dovich effect) and secondary cosmological effects (e.g. integrated Sachs-Wolfe effect, Rees-Sciama effect). 

Any signal that has a different spectral signature than the primordial CMB will be removed during the LGMCA processing, i.e. Galactic foregrounds and the thermal Sunyaev-Zel'dovich effect. Achromatic effects (e.g., kinetic Doppler, kSZ, ISW) will not be removed by the LGMCA method. If there are physical models for these remnant effects, these can be modeled and thereafter subtracted as we do hereafter for the kDq and kSZ effects (Section \ref{sec:astro}) and for the ISW effect (Section \ref{sec:ISW}).

\label{sec:datapr1}

\subsection{Astrophysical secondary effects}\label{sec:astro}

\subsubsection{The kinetic Doppler quadrupole (kDq)}

The large scale coherent motion of our galaxy within the local group, and of the local group of galaxies with respect to the CMB frame, create a large Doppler dipole in the CMB observations, but also a smaller quadrupole effect. The kDq contribution is given by: 
\begin{equation} 
\delta_{kD,\ell=2}=\left(\frac{v}{c}\right)^2\left[cos^2 \theta-\frac{1}{3}\right],
\end{equation} 
where $\theta$ is the angle between the position on the sky and the direction of motion creating the kinetic Doppler quadrupole \citep{Copi:2005ff}. 
We have used the kDq map freely available from \cite*{Rassat:2012}\footnote{the kinetic Doppler quadrupole map is downloadable from \url{http://www.cosmostat.org/anomaliesCMB.html}}.

In  \cite{PlanckStat}, the kDq was subtracted after inpainting, and it is not clear if the Wiener inpainting which assumes isotropy and Gaussianity can properly reconstruct the missing part relative to the kDq,
which is obviously not Gaussian. This could be one of the reasons why  the Wiener method impacts anomalies such as alignments, as observed in \cite{Copi2013alignements}. Here we subtract the kDq effect after sparse inpainting.
 
\subsubsection{The kinetic Sunyaev-Zel'dovich (kSZ) effect}

\begin{figure}[htbp]
   \centering
   \includegraphics[width=8cm]{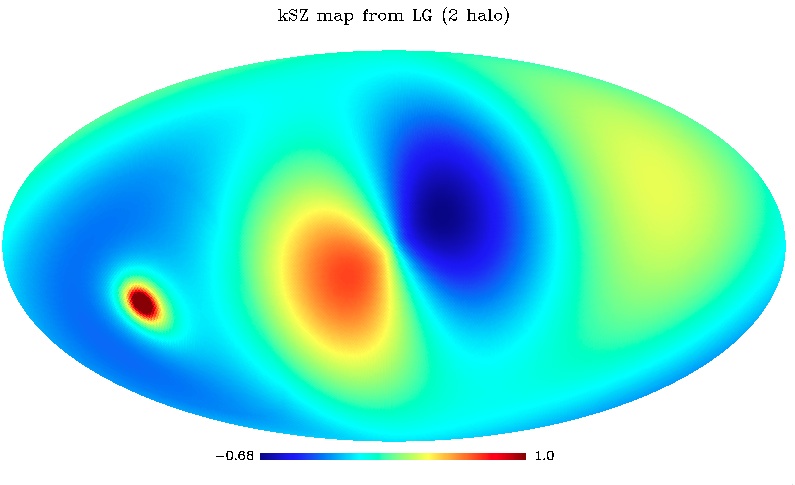} 
   \caption{Kinetic Sunyaev-Zel'dovich ($\mu K$) effect due to the local group of galaxies, assuming two haloes centred on Andromeda and the Milky Way. The maximum value in the figure was reduced to $1\mu K$ (instead of $4.1\mu K$) so that the kSZ structure due to the Milky Way (in the centre of the image) can be seen by eye.}
   \label{fig:ksz}
\end{figure}

\begin{figure*}[htbp]
   \includegraphics[width=7.7cm]{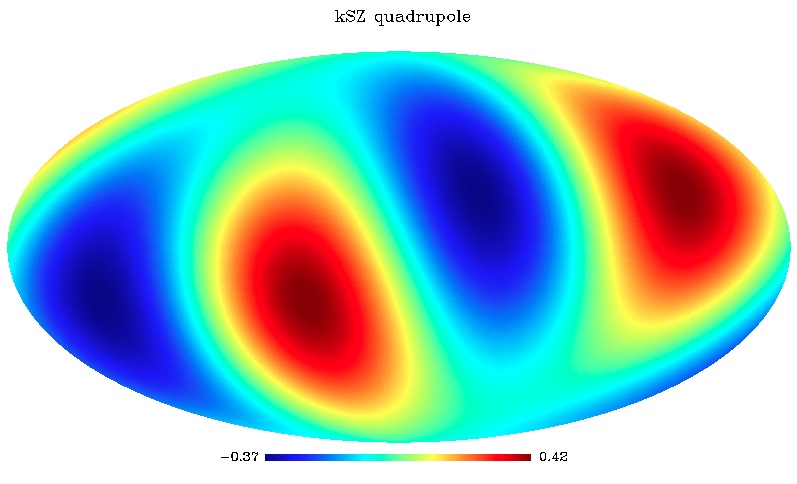}\includegraphics[width=7.7cm]{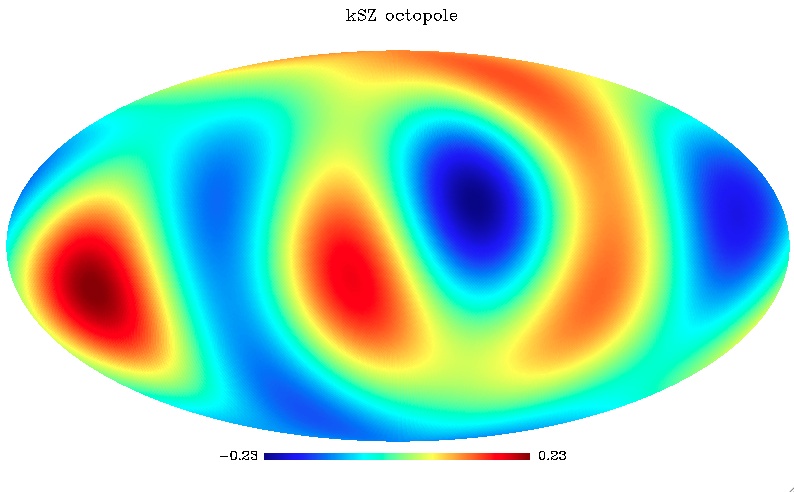} 
   \caption{Quadrupole (\emph{left}) and octopole (\emph{right}) of the kinetic Sunyaev-Zel'dovich (in $\mu K$) effect due to the local group of galaxies, assuming two haloes centred on Andromeda and the Milky Way. The map presents an aligned quadrupole and octopole which is due to the halo of the Milky Way.}
   \label{fig:kszquadoct}
\end{figure*}

The motion of our local group with respect to the cosmic microwave background produces another distortion to the observed CMB photos called the kinetic Sunyaev-Zel'dovich effect (kSZ). This is caused by CMB photons scattering off \emph{moving} free electrons in the gaseous halo of the local group. The kSZ effect introduces an achromatic thermal shift, and will therefore not be removed through the LGMCA component separation technique. The thermal shift is given by: 
\begin{equation} \frac{\Delta T}{T}({\bf {\hat n}}) = -\frac{1}{c}\left({\bf v}_{\rm LG-CMB}\cdot {\bf \hat{r}}\right)\tau({\bf \hat{n}})\end{equation}
where $\tau({\bf \hat{n}})$ corresponds to the optical depth along the line of sight \citep{Rubin:ksz}, and ${\bf v}_{\rm LG-CMB}$ is the velocity of the local group with respect to the rest frame of the CMB.  We calculate the optical depth map $\tau({\bf \hat{n}})$ using the same method as in \cite{Rubin:ksz}, except we assume separate haloes for both Andromeda and the Milky Way. The kSZ effect resulting from the Milky Way is interesting because it presents an aligned quadrupole and octopole (see Figures \ref{fig:ksz} and \ref{fig:kszquadoct}), however because its amplitude is very low, subtracting it from the observed CMB should have little effect on the anomalies. 

\subsection{Cosmological secondary effects: the integrated Sachs-Wolfe effect}\label{sec:ISW}
On much larger scales, CMB photons will also travel through the large-scale gravitational potential as they travel from the surface of last scattering to us. On infall into the potential they gain energy, which they lose upon exit if there is no change in the large-scale cosmic gravitational potential, as is the case in a flat universe with no dark energy, since in this case growth of structure will exactly counter-balance the effect due to the expansion of the universe. However, for universes with dark energy, curvature or modified gravity models, the photons may gain or lose energy, which will introduce a thermal shift in the CMB photons.  

\begin{figure}[htbp]
   \centering
   \includegraphics[width=8.5cm]{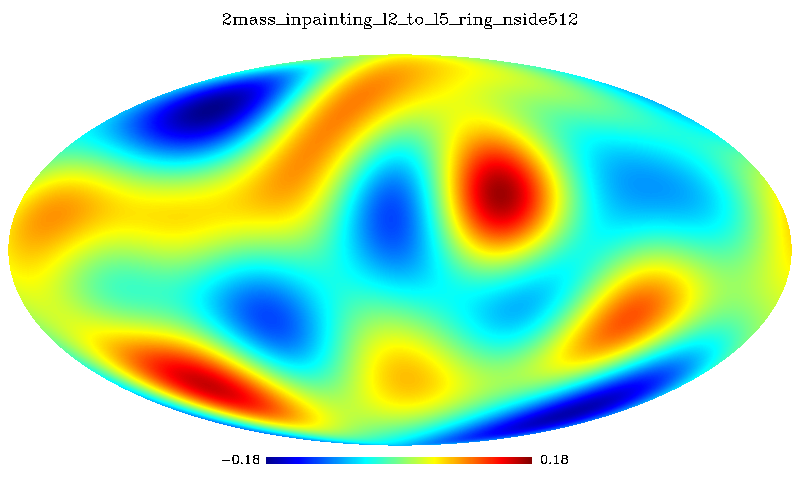}\includegraphics[width=8.5cm]{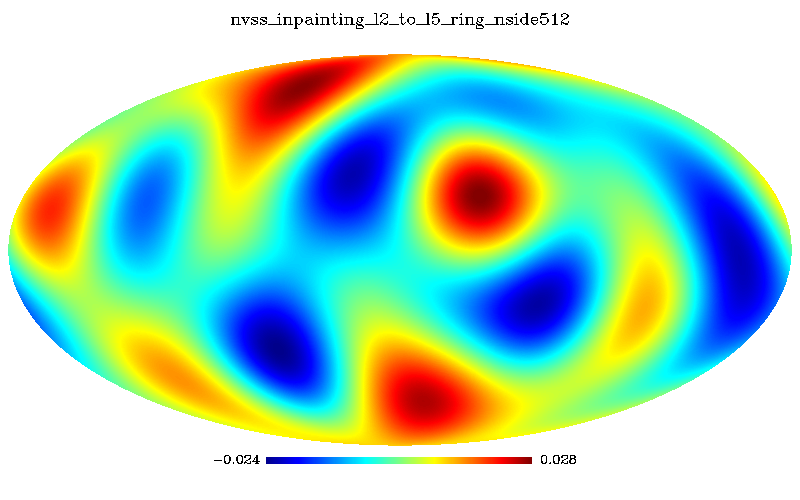}
   \caption{\emph{Left:} Galaxy density map for 2MASS galaxies and \emph{right}: for NVSS. The maps are reconstructed along the Galactic plane using sparse inpainting up to $\ell=64$ using $nside=512$ and shown here for $\ell=2-5$. The maps correspond to the $g_{\ell m}$ coefficients in Equation \ref{eq:isw} for $\ell=2-5$ and $m \in [-\ell,\ell]$.}
   \label{fig:tracers}
\end{figure}

\begin{figure}[htbp]
   \centering
   	\hspace{-1cm}\includegraphics[width=9.5cm]{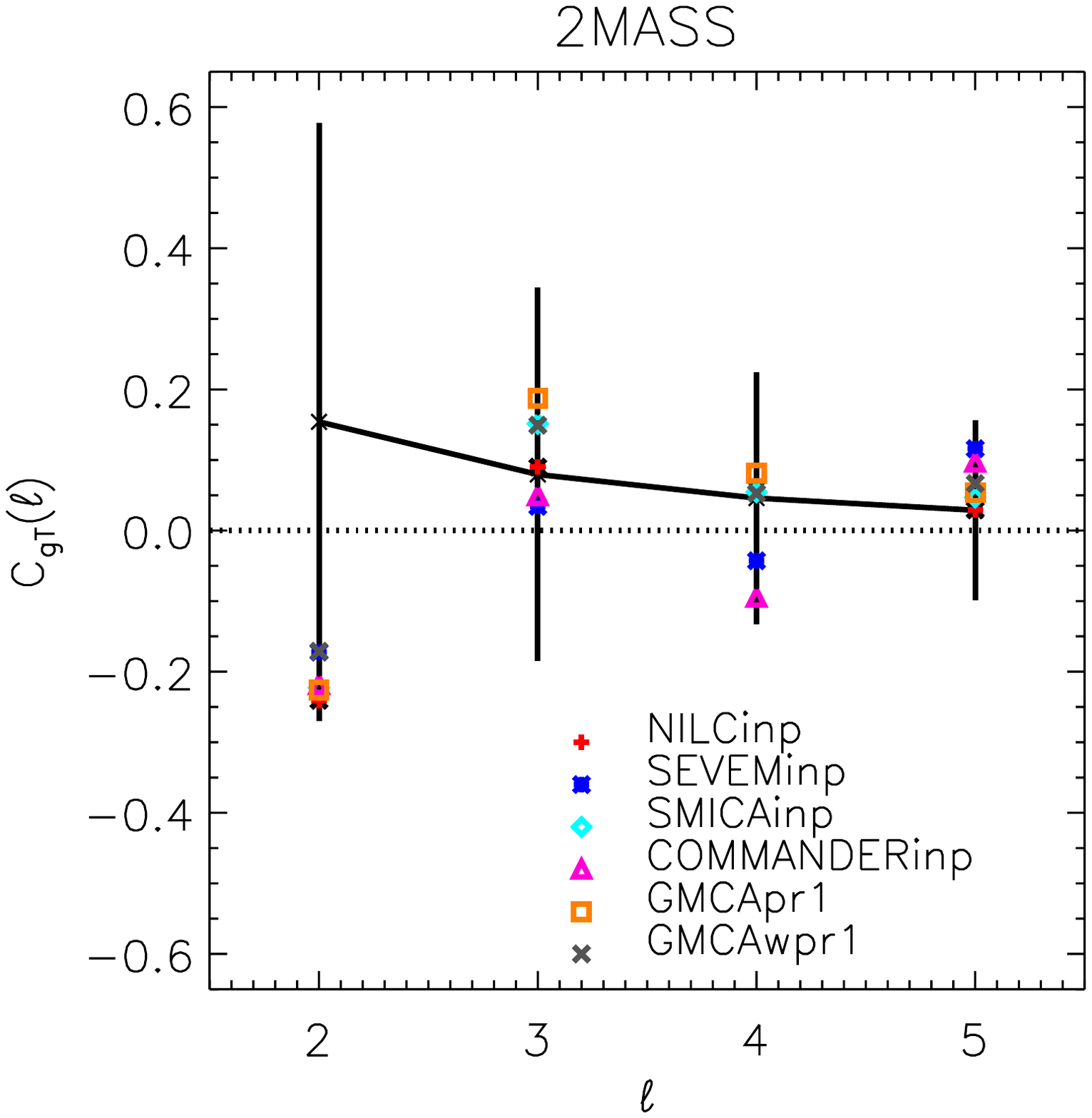}\hspace{-2.8cm}\includegraphics[width=9.5cm]{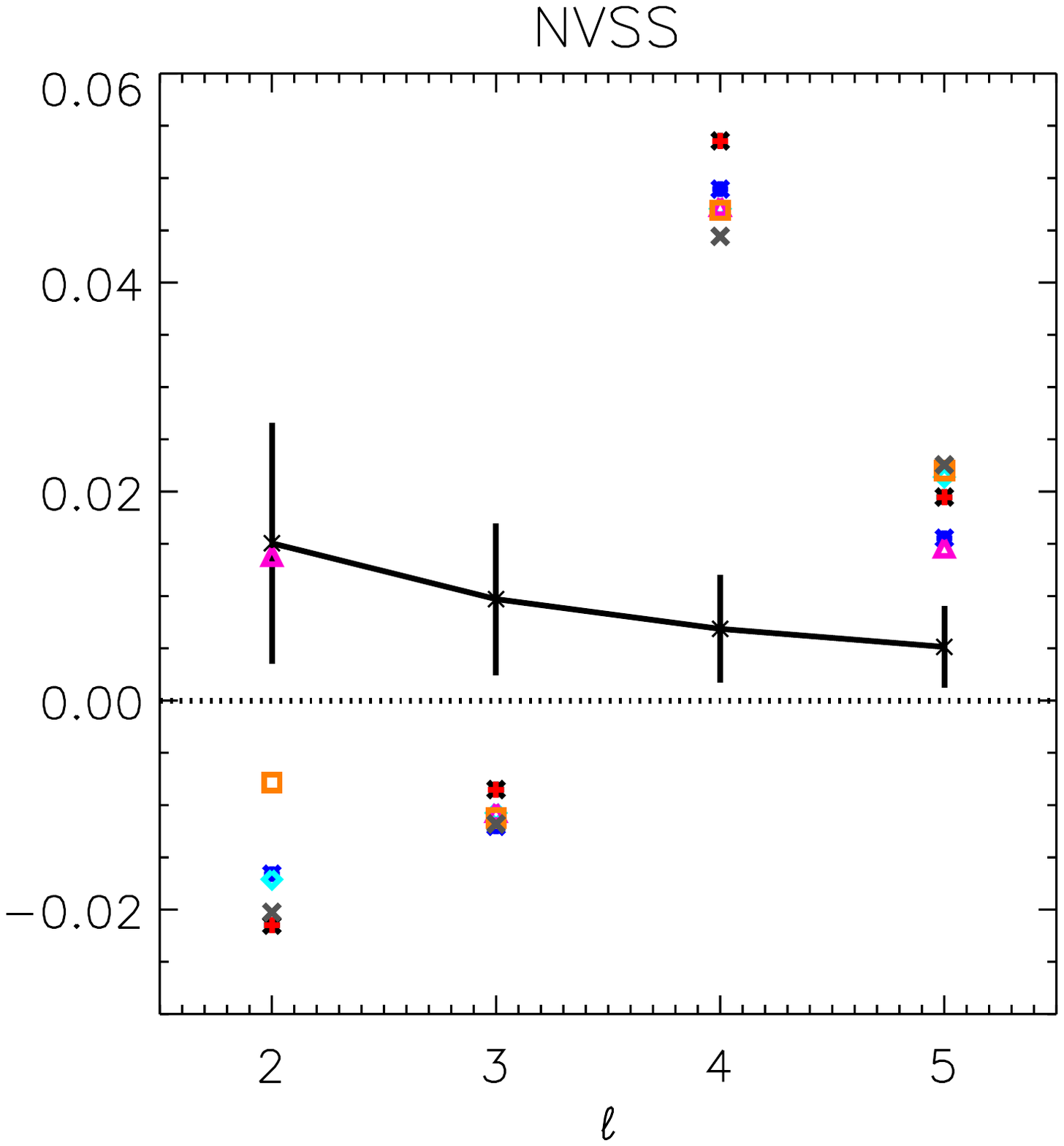}
	\vspace{-3cm}
   \caption{Cross-correlation $C_{gT}(\ell)$ of various CMB maps with 2MASS (\emph{left}) and NVSS (\emph{right}), in units of $\mu K$. The theoretical prediction is shown in solid black, along with error bars \citep[see for e.g.,][]{Rassat:2007krl} which are dominated by cosmic variance. In the error bar calculation, we have taken $f_{\rm sky}=0.69$ for 2MASS and $f_{\rm sky}=0.66$ for NVSS.}
   \label{fig:cgt}
\end{figure}

\begin{figure}[htbp]
   \centering
   	\hspace{-1cm}\includegraphics[width=9.5cm]{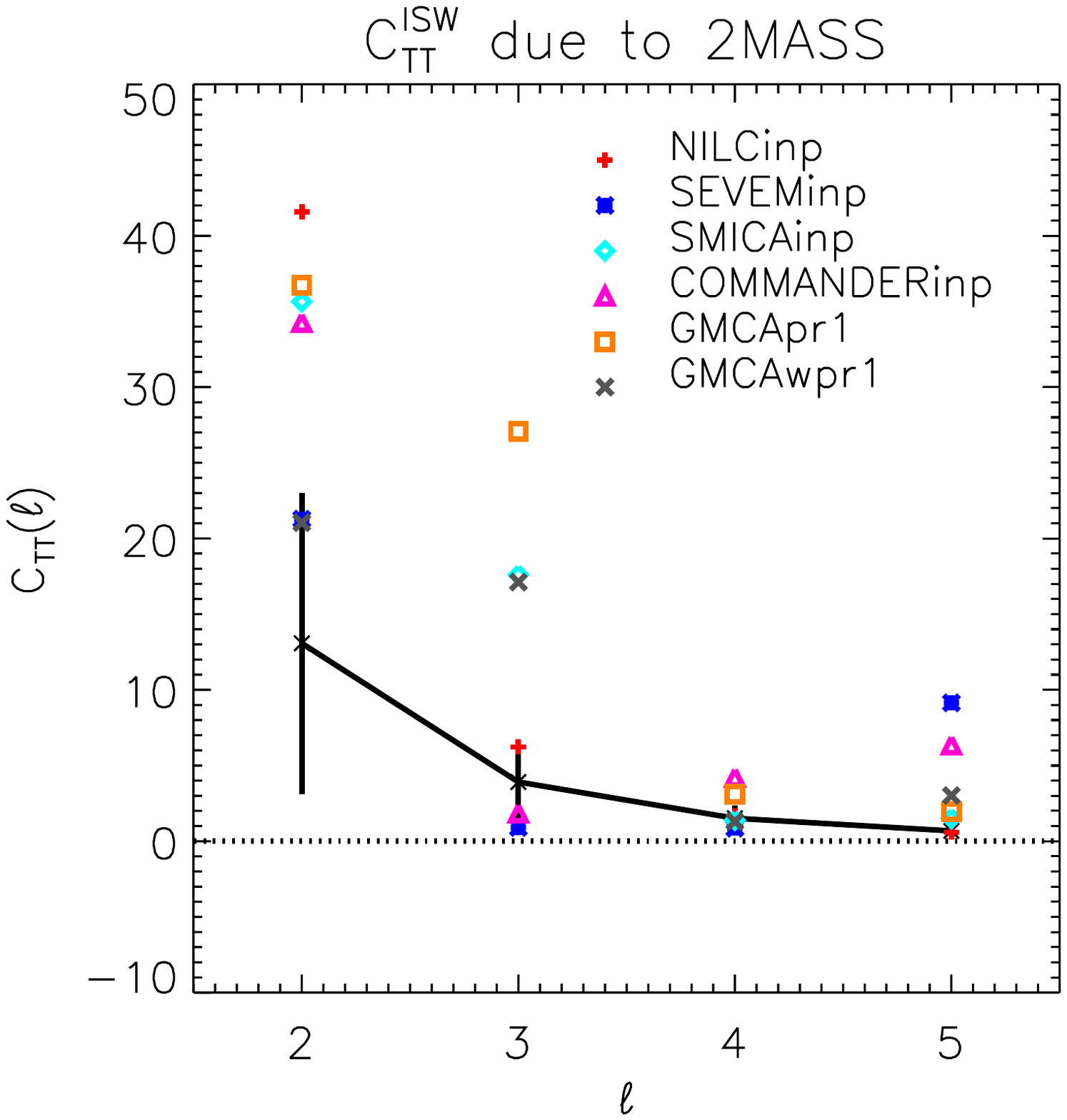}\hspace{-3.cm}\includegraphics[width=9.5cm]{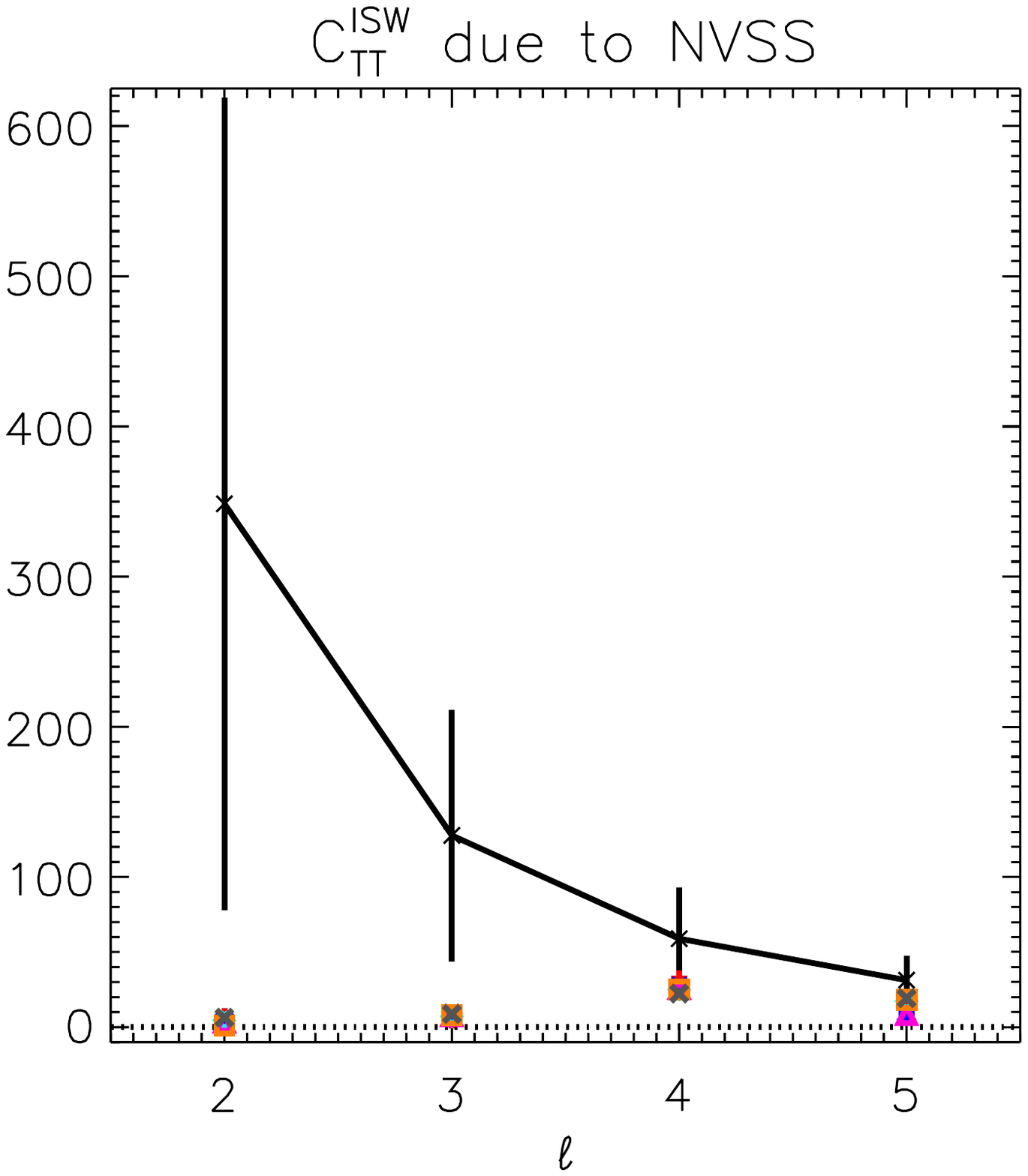}
	\vspace{-3cm}
   \caption{Temperature anisotropy power of the ISW effect ($C_{TT}^{ISW}$) due to 2MASS (\emph{left}) and NVSS (\emph{right}) in units of $\mu K^2$. Since the signal is determined using the cross-correlation from the data, the signal can vary for different CMB map renditions. The theoretical prediction is shown in solid black, along with error bars which are dominated by cosmic variance. In the error bar calculation, we have taken $f_{\rm sky}=0.69$ for 2MASS and $f_{\rm sky}=0.66$ for NVSS.}
      \label{fig:ctt}
\end{figure}

\begin{figure}[htbp]
   \centering
   	\hspace{-1cm}\includegraphics[width=9.5cm]{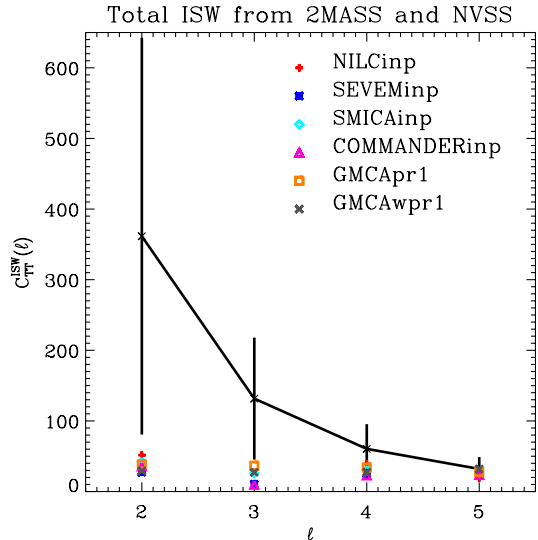}
		\vspace{-3cm}
   \caption{Temperature anisotropy power of the ISW effect ($C_{TT}^{ISW}$) due to 2MASS and NVSS (\emph{right}) in units of $\mu K$. Since the signal is determined using the cross-correlation from the data, the signal can vary for different CMB map renditions.}
   \label{fig:ctttotal}
\end{figure}

\begin{figure}[htbp]
   \centering
   \includegraphics[width=8.5cm]{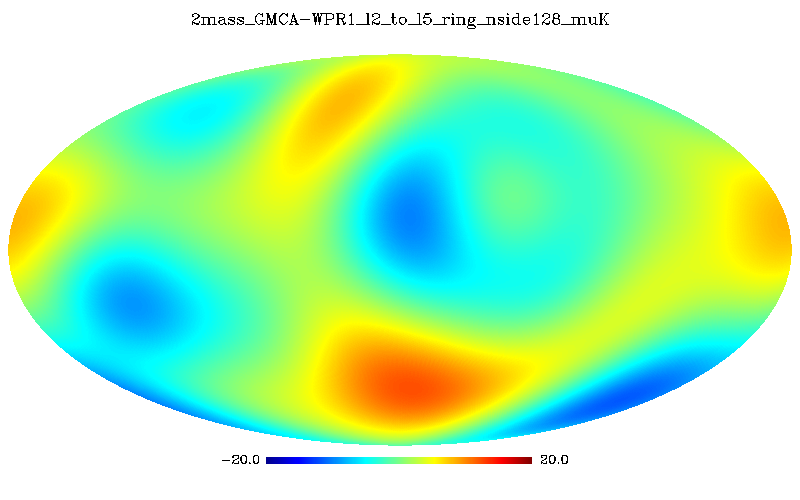}\includegraphics[width=8.5cm]{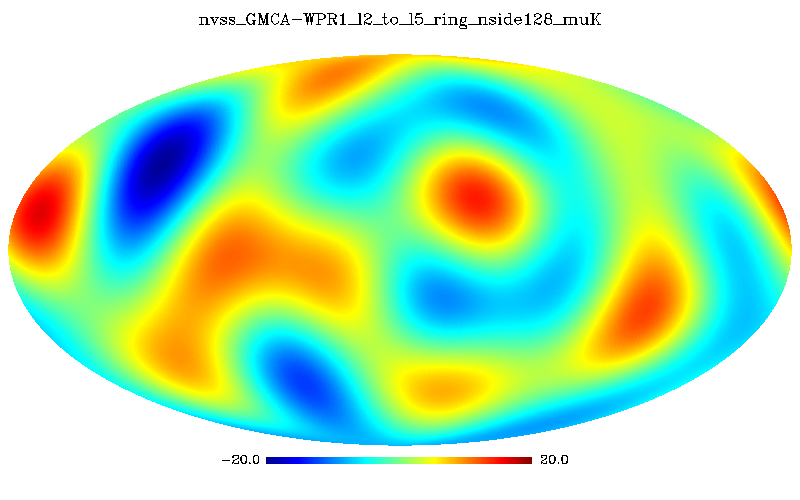}
   \caption{Reconstructed ISW field due to 2MASS galaxies (\emph{left}) and due to NVSS (\emph{right}) for $\ell=2-5$. The amplitude of the ISW signal is determined using the cross-correlation from the data, here using the LGMCA-WPR1 rendition of the CMB map.}
   \label{fig:iswmap}
\end{figure}

\begin{figure}[htbp]
   \centering
   \includegraphics[width=8.5cm]{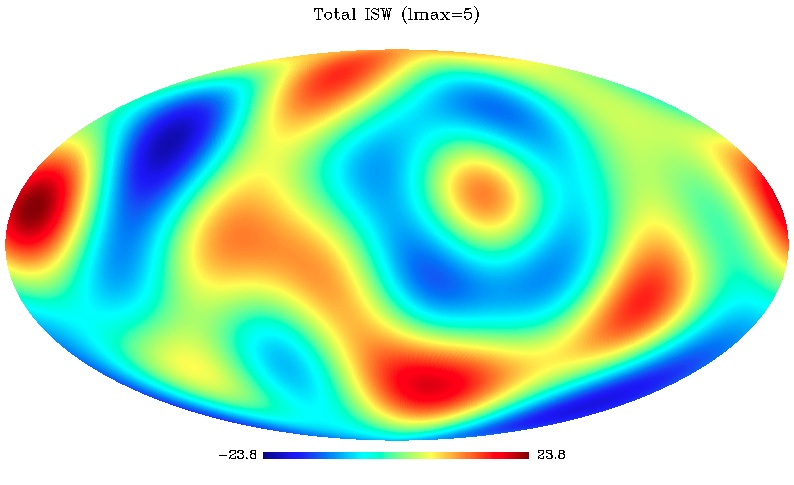}\includegraphics[width=8.5cm]{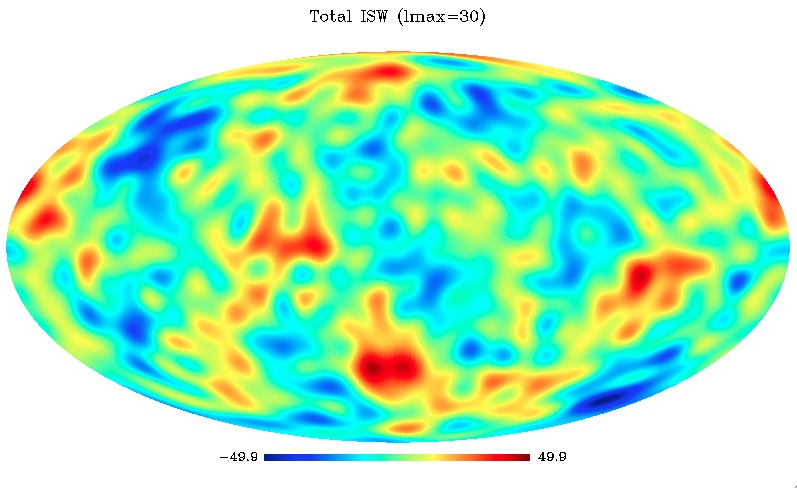}
   \caption{Reconstructed ISW field due to both 2MASS galaxies and NVSS for $\ell=2-5$ (\emph{left}) and $\ell=2-30$ (\emph{right}). The amplitude of the ISW signal is determined using the cross-correlation from the data, here using the LGMCA-WPR1 rendition of the CMB map.}
   \label{fig:iswmap:total}
\end{figure}

Following \cite{Rassat:2012}, we estimate the ISW signal from full-sky CMB and large scale structure (LSS) maps, using:  \begin{equation} \delta^{\rm ISW}_{\ell m} = \frac{C_{gT}(\ell)}{C_{gg}(\ell)}g_{\ell m},\label{eq:isw}\end{equation} where $C_{gT}(\ell)$ and $C_{gg}(\ell)$ are the galaxy-temperature cross-correlation and galaxy auto-correlation respectively. These can be measured directly from the data in order to be as model independent as possible (though Equation \ref{eq:isw} is not entirely model independent).  We note that Equation \ref{eq:isw} is also independent of the galaxy bias.  Figure \ref{fig:tracers} shows the reconstructed density maps for 2MASS (\emph{left}) and NVSS (\emph{right}), i.e. $g_{\ell m}$ for $\ell=2-5$ after inpainting has been applied as described in \cite{Rassat:2012,Rassat:axes} with $nside=512$.

In practice, we estimate the full-sky local ISW signal from 2MASS \citep{Jarrett:2004a} and NVSS \citep{NVSS} surveys in the same way as in \cite*{Rassat:2012} and \cite{Rassat:axes}. The ISW signal is estimated from the data, meaning that for each CMB map the ISW amplitude can vary (from the $C_{gT}$ term in Equation \ref{eq:isw}). The amplitude of $C_{gT}(\ell)$ for each CMB map is shown in Figure \ref{fig:cgt} for 2MASS (\emph{left}) and NVSS (\emph{right}). The theoretical prediction is shown in solid black, along with error bars \citep[see for e.g.,][]{Rassat:2007krl} which are dominated by cosmic variance, where we have taken $f_{\rm sky}=0.69$ for 2MASS and $f_{\rm sky}=0.66$ for NVSS for the error bar calculation. The observed 2MASS cross-correlations is always within the $1\sigma$ error bars, however we note that the NVSS cross-correlation is not in agreement with the theoretical predictions, as was already reported in \cite{Rassat:2012} and \cite{Rassat:axes}

In Figure \ref{fig:ctt} we show the reconstructed temperature anisotropy power $C_{TT}^{ISW}(\ell)$ due to either 2MASS or NVSS galaxies, which can be estimated by: 
\begin{equation}C^{ISW}_{TT}(\ell) = \frac{C_{gT}^2(\ell)}{C_{gg}(\ell)}.\end{equation} We find with Planck data, the amplitude of the signal is similar for both 2MASS and NVSS data to what we had found in \cite{Rassat:2012} and \cite{Rassat:axes} with WMAP data.
The total power of the ISW effect for both 2MASS and NVSS is shown in Figure \ref{fig:ctttotal} and the descrepency with the theoretical prediction is mostly due to the low signal arising from NVSS. In Figure \ref{fig:iswmap}, we show the corresponding temperature anisotropy maps of the ISW signal due to 2MASS (\emph{left}) and NVSS (\emph{right}), and in Figure \ref{fig:iswmap:total} the summed contribution due to both surveys for $\ell=2-5$ (\emph{left}) as well as for $\ell=2-30$ (\emph{right}) shown for illustration, since only the largest scales ($\ell=2-5$) are used in this analysis.

\section{Results}
\subsection{CMB Data}

We use six different CMB maps to study the large scale anomalies in the CMB, as summarised in Table \ref{tab:data}. Four come from the official PR1 Planck release: Nilc, Sevem, Smica, and Commander with respective $f_{\rm sky}$ values of 0.92, 0.76, 0.96 and 0.75.\footnote{The official PR1 maps were taken downloaded from \url{http://pla.esac.esa.int/pla/aio/planckResults.jsp?}}
The Nilc and Smica have been updated since the first release in March 2013, and we have used the last versions available (R1.20).  It is not clear however if these versions correspond to the maps analysed in \cite{PlanckStat}, so a comparison with published results is difficult. 
We inpaint these four maps using sparse inpainting (see Section \ref{sec:processing}) and reconstruct the harmonic coefficients up to $\ell=64$. These results can be reproduced using the  following command line in the open source sparse inpainting package {\tt ISAP} software\textcolor{red} {\footnotemark[1]}:
\vspace{0.1cm}

{\tt > alm = cmb\_lowl\_alm\_inpainting(map, Mask, lmax=64, niter=100, InpMap=result)}\vspace{0.1cm}

\begin{figure}[htbp]
    \centering
    \includegraphics[width=8.5cm]{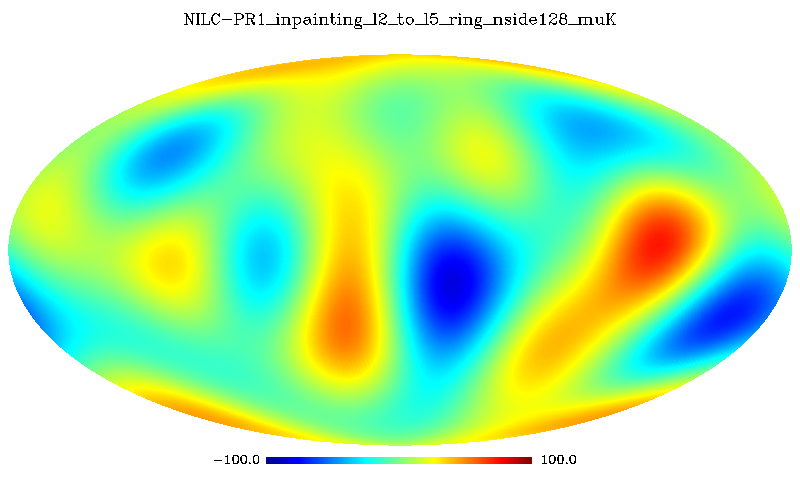}\includegraphics[width=8.5cm]{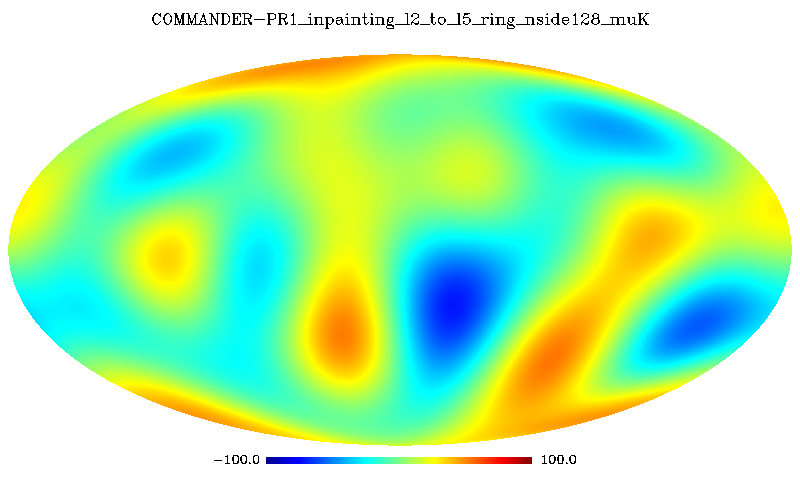}
    
   \includegraphics[width=8.5cm]{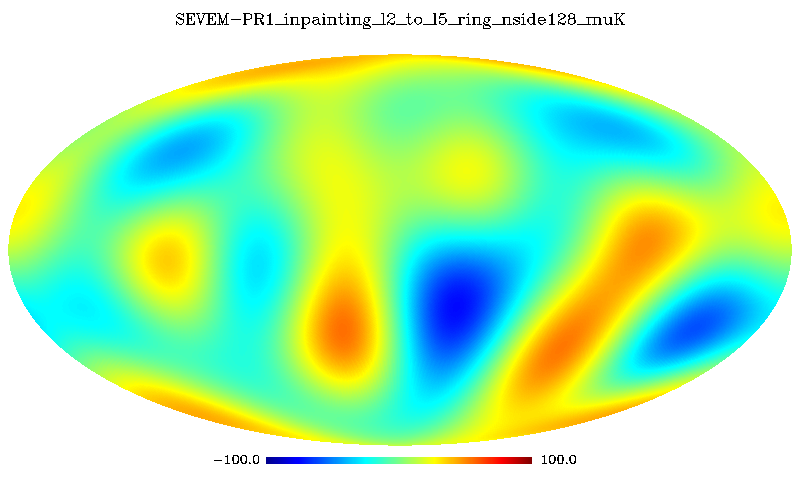}\includegraphics[width=8.5cm]{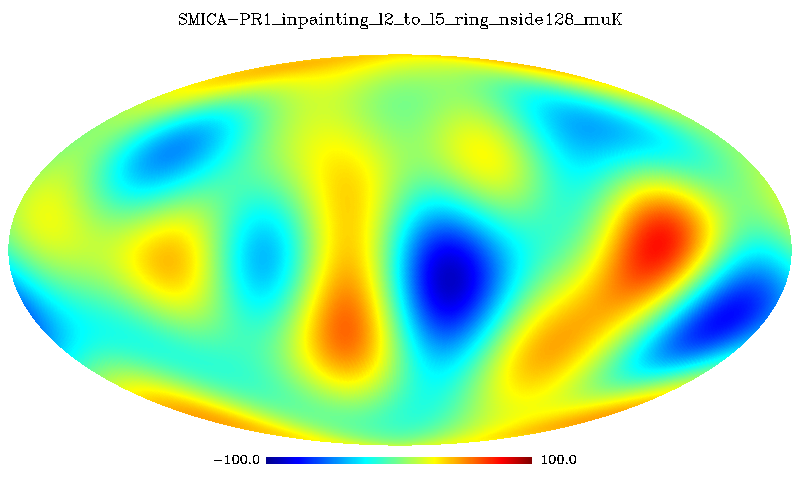}
        \caption{Difference between the four official Planck maps after inpainting for their respective maps (clockwise: Nilc, Commander, Smica and Sevem) and the LGMCA map, for $\ell=2-5$, in $\mu K$.}
    \label{fig:planckvslgmca}
 \end{figure}

Figure \ref{fig:planckvslgmca} shows the difference between the official Planck maps after inpainting for their respective maps (clockwise: Nilc, Commander, Smica and Sevem) and the LGMCA map, for $\ell=2-5$

We consider two other maps, PR1-LGMCA and WPR1-LGMCA, obtained using 
the LGMCA component separation method respectively on Planck-PR1 data and on WMAP 9yr and Planck-PR1 data jointly.
These two maps are full-sky and do not require any additional processing. In the spirit of reproducible research, the maps, codes and scripts are freely available at \url{http://www.cosmostat.org/planck_wpr1.html}. 

\begin{figure}[htbp]
   \centering
\includegraphics[width=8.5cm]{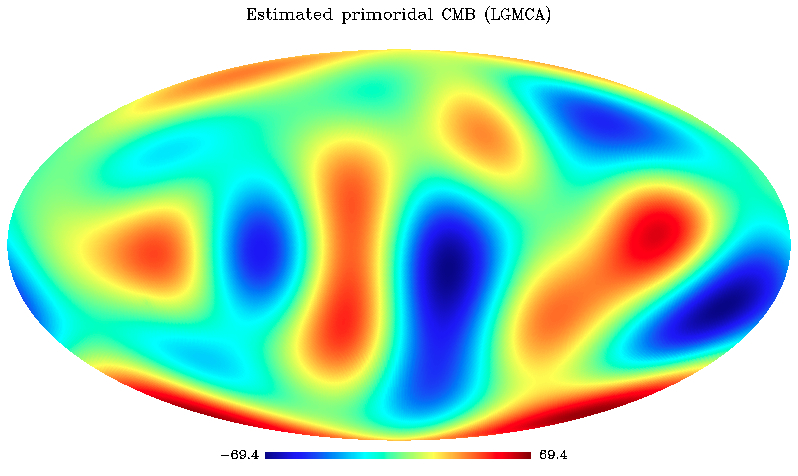}\includegraphics[width=8.5cm]{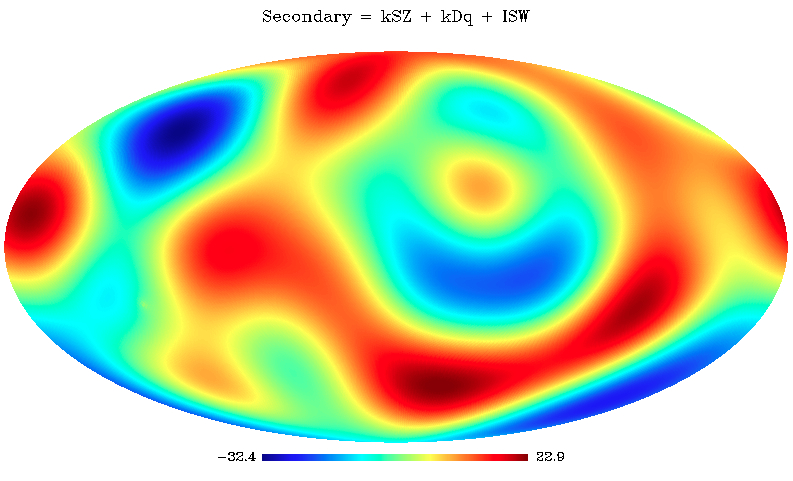}
\includegraphics[width=8.5cm]{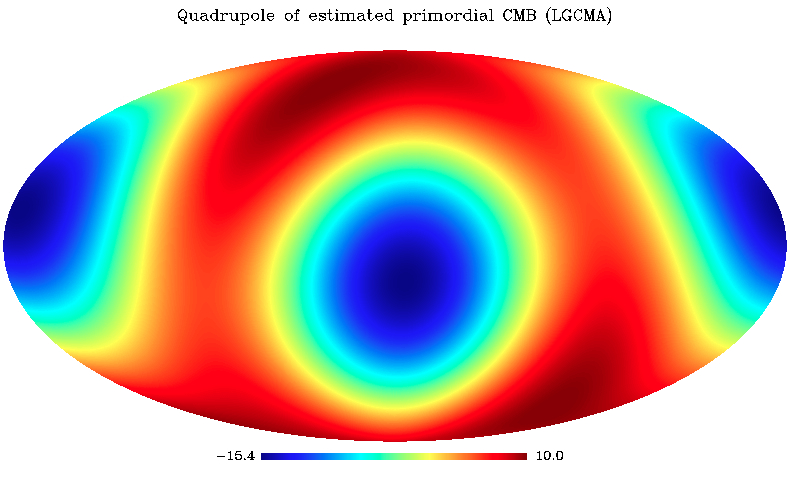}\includegraphics[width=8.5cm]{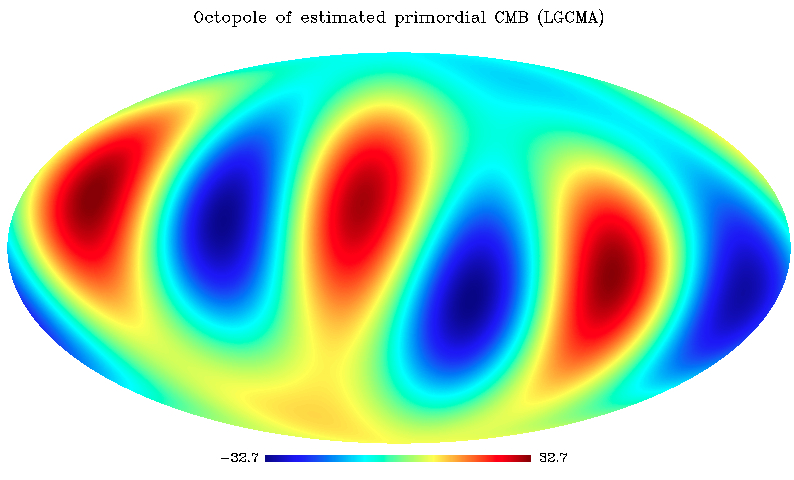}
\includegraphics[width=8.5cm]{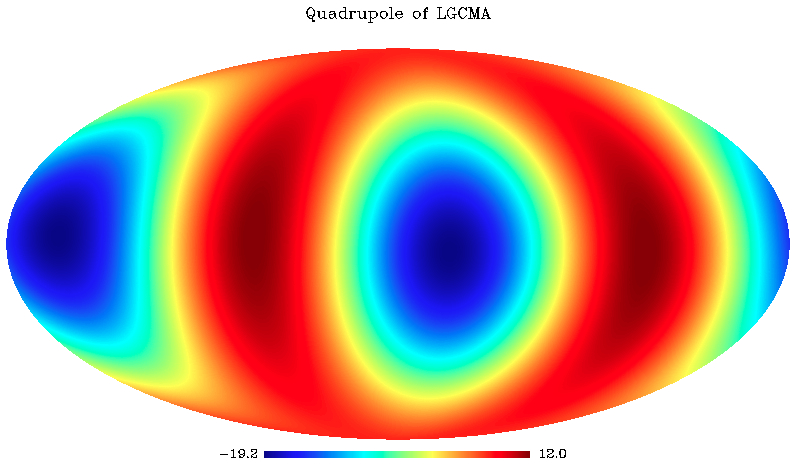}\includegraphics[width=8.5cm]{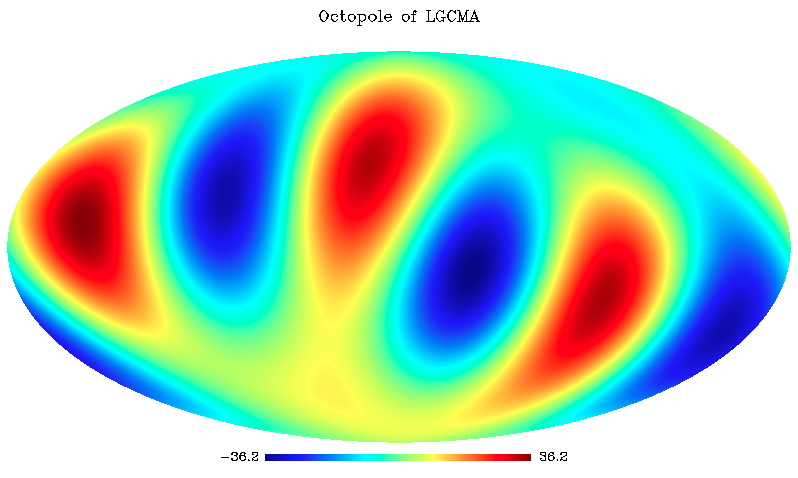}
   \caption{\emph{Top left}: Estimated primordial CMB map for $\ell=2-5$ using the LGMCA map and where secondary astrophysical and cosmological signals considered in this paper have been removed. \emph{Top right}: The total contribution of secondary astrophysical and cosmological signals due to the kinetic Sunyaev-Zel'dovich, kinetic Doppler quadrupole and ISW signal estimated from 2MASS and NVSS surveys. The amplitude of the ISW map is estimated from the data, using the LGMCA-WPR1 map for the CMB data. \emph{Middle}: Quadrupole (\emph{left}) and octopole (\emph{right}) of the estimated primordial CMB for the LGMCA map, compared with the quadrupole (\emph{bottom left}) and octopole (\emph{bottom right}) of the LGMCA map before secondary signals are subtracted. }
   \label{fig:lgmca_secondary}
\end{figure}

Figure \ref{fig:lgmca_secondary} shows the LGMCA-WPR1 map after subtraction of the secondary astrophysical and cosmological signals (\emph{top left}), as well as the summed temperature fields of these signals (\emph{top right}). The isolated quadrupole and octopole of the estimated primordial LGMCA map are shown in the middle, and can be compared with the quadrupole and octopole of the LGMCA before subtraction of secondary signals.

\begin{table}[htbp]
   \centering
   \begin{tabular}{@{} lccc @{}} 
Name & Data & $f_{sky}$&Additional\\
&&& processing\\
\hline
\hline
Nilc&PR1 Planck&$0.92$& inpainted\\
Sevem& PR1 Planck&$0.76$&inpainted\\
Smica& PR1 Planck&$0.96$&inpainted\\
Commander& PR1 Planck&$0.75$&inpainted\\
PR1-LGMCA&PR1 Planck& full-sky&-\\
\multirow{2}{*}{WPR1-LGMCA}&PR1 Planck&\multirow{2}{*}{full-sky}&\multirow{2}{*}{-}\\
&and W9&&\\
\hline
   \end{tabular}
   \caption{Overview of CMB maps used in the analysis of this paper, with corresponding masks and mask processing details. The first four maps correspond to the official Planck PR1 maps with their respective masks}. 
   \label{tab:data}
\end{table}

\subsection{Analysis}
In tables \ref{tab:quad} to \ref{tab:parity}, we report the results for the low quadrupole (\ref{tab:quad}), the quadrupole-octopole alignment (\ref{tab:quadoct}), the planar octopole (\ref{tab:planar}), the AoE (\ref{tab:aoe}) and mirror parity (\ref{tab:parity}). Our aim is first to compare if the reported anomalies are still existent in the various official Planck renditions as compared to the LGMCA renditions considered here, in order to assess the possible influence of the masking process. We are then interested to see if the reported anomalies still persist after subtraction of the different astrophysical and cosmological secondary effects (kDq, kSZ, ISW). 

As reported in \cite{PlanckStat}, we find that the quadrupole-octopole alignment is significant in several official Planck renditions (Sevem, Nilc and Smica, but not in Commander as was reported by the Planck team). However, after subtraction of the kDq, all maps (including the LGMCA maps but except Commander) have an even more anomalous quadrupole/octopole alignment, with the corresponding probabilities from simulations ranging from $0.02-2.80\%$ (see Table \ref{tab:quadoct}, and similarly to what \cite{Schwarz:2004} had found with WMAP data, \cite{Copi2013alignements} with Planck data, and what \cite{PlanckStat} found even though the actual values are slightly different). However, after subtraction of the ISW  effect, this alignment is no longer significant. Further subtraction of the kSZ effect does not affect the alignment significance. 

For all other anomalies: the planar octopole, the AoE, positive and negative mirror parity, we find that none of the statistics considered here are anomalous, whether considering the official Planck maps, the LGMCA renditions, and whether any of the astrophysical and cosmological secondary effects considered here are removed or not (see Tables \ref{tab:planar}-\ref{tab:parity}), except for two exceptions. Firstly, (Smica-kDq) returns  a significant Axis of Evil, with a mean interangle of only 21$^\circ$ (with only $0.06\%$ of simulations returning a similar value). The AoE persists after ISW subtraction, but does not persist after further subtraction of the kSZ map. Secondly, (Commander-kDq)  returns a negative mirror parity at the $\sim 2\sigma$ level (with $3.4\%$ of simulations returning such a high value for $S_-$), which persists both after subtraction of the ISW and after subtraction of the kSZ effect. We note that \cite{PlanckStat} had reported an anomalous positive mirror parity, which we do not report. Regarding the AoE, we note that the LGMCA-WPR1 map (after kDq/ISW and kDq/ISW/kSZ subtraction) presents an usually large mean interangle ($\sim 70^\circ$), and that only $3.8-4.4\%$ of simulations present such a large value. 

Regarding the low quadrupole, we find that nearly all maps return somewhat anomalous values of the low quadrupole at the $< 2 \sigma$ level (i.e. $1.4-3.9\%$, except PR1-LGMCA for which $5.8\%$ of simulations return similar values of the quadrupole), and that subtraction of the kDq makes the anomaly slightly more significant ($1.2-3.4\%$, except for PR1-LGMCA, where only $4.6\%$ of simulations return a similar value, i.e. just above the $2\sigma$ level).  For all maps, the quadrupole power in decreased after subtraction of the ISW signal, however the expected theoretical signal is also reduced by the expected amount of ISW signal. These are two aspects of assessing the low quadrupole anomaly and the final probability is related to both of these effects, and a lower quadrupole after subtraction, does not necessarily mean a more anomalous quadrupole. In Table 3, we find that these two effects counter-balance, such that subtraction of the ISW effect (and subsequently of the kSZ effect) does not alter the significance of the anomaly by much, though for PR1-LGMCA, it becomes slightly less significant (from $4.6\%$ to $5.0\%$).

 \section{Conclusions}
In this paper, we investigate possible sources of the reported anomalies in the Cosmic Microwave Background. We focus on three possible issues: mask processing, astrophysical secondary effects (kDq and kSZ effects) and cosmological secondary effects (the ISW effect). 

Our first conclusion is that using full-sky data has less impact on the CMB statistics than including mask processing to limit systematics due to foreground residuals. This striking result is only possible thanks to i) the new Planck data set which combined with WMAP gives give 14 frequency channels and ii) the development of new component separation methods such as LGMCA which provide high quality full-sky maps.  One remarkable aspect of the impact of mask processing is that more conservative masks naturally increase the significance of the cold spot anomaly. With the full-sky LGMCA maps we find that the kurtosis value of the wavelet scale related to the size of the cold spot anomaly is no longer significant. 

In our study we therefore use two full-sky LGMCA maps, one derived from Planck data alone, and another derived from both Planck and WMAP maps, as well as the four official Planck PR1 maps (Nilc, Sevem, Smica and Commander) to which we apply sparse inpainting. 

We study these maps focussing on several reported anomalies: the low quadrupole, the quadrupole/octopole anomaly, the planar octopole, the Axis of Evil, positive/negative mirror parity and the cold spot. 

For the secondary effects, we consider the kinetic Doppler effect \citep[using the publicly available map released in ][]{Rassat:2012}, the integrated Sachs-Wolfe effect from 2MASS and NVSS galaxies \citep[using updated versions of the public maps from][]{Rassat:2012,Rassat:axes}, and create a new kinetic Sunyaev-Zel'dovich map using the method put forward by \cite{Rubin:ksz} but modified to include separate haloes for Andromeda and the Milky Way. The resulting kSZ presents an intrinsic quadrupole/octopole alignment which is interesting since it is similar to that observed in the CMB. However, the amplitude of the kSZ effect is too low for its subtraction to have an impact on any of the anomalies considered here.

With these astrophysical and cosmological secondary effects reconstructed, we can subtract them from the observed CMB in order to create an estimate of the primordial CMB and test it for the reported anomalies. 

Our conclusions regarding the claimed anomalies are summarised in Table \ref{tab:summary}. We find that the octopole planarity, AoE, mirror parity and cold spot are never anomalous, whether after kDq subtraction or after subsequent subtraction of the ISW and kSZ effects (with two exceptions regarding the Smica and Commander maps, see Table \ref{tab:summary}). On the contrary, we find that after subtraction of the kDq effect, the quadrupole/octopole alignment is still anomalous (except for the Commander map). However, after subsequent subtraction of the ISW and kSZ maps, the alignment is no longer significant. Regarding the low quadrupole, we find that nearly all maps return significantly low values, whether any secondary effect has been subtracted or not, similarly to what \cite{Rassat:2012} had found with WMAP data. We note that the significance of only one anomaly (the quadrupole/octopole alignment) is affected by subtraction of secondary effects out of six anomalies considered.

 \begin{table*}[htbp]
   \centering
   \begin{tabular}{@{} lccccc @{}} 
   \hline
Anomaly &   Inpainted or full-sky \& kDq sub.&kDq, ISW, kSZ sub.  \\
\hline
Low quad  &   anomalous &  as anomalous \\
Quad/oct alignment  &  anomalous& not anomalous\\
	&(except Commander)&\\
Oct planarity & not anomalous & not anomalous\\
Axis of Evil &not anomalous &not anomalous\\
&(except Smica)&\\
Odd mirror parity &not anomalous&not anomalous\\
&(except Commander&(except Commander\\
& at $2\sigma$ level)&at $2\sigma$ level)\\
Even mirror parity &not anomalous&not anomalous\\
Cold spot  &  not anomalous   &  -\\
\hline
   \end{tabular}
   \caption{Summary of conclusions in this paper for the various anomalies considered.  The first column lists the anomaly considered. The second shows the overall conclusion of the significance for the six maps considered in this paper (which are either official Planck maps that are inpainted or full-sky LGMCA maps), for which the kDq map has been subtracted. The final column shows the overall conclusion of the significance of the six maps after the ISW and kSZ have further been subtracted.} 
   \label{tab:summary}
\end{table*}

 \acknowledgments
The authors thank Dragan Huterer, Glenn Starkman and Craig Copi for stimulating discussions regarding the large-scale anomalies and their statistics. AR thanks Matthew Nichols and Douglas Rubin for useful discussions about the local group and the kinetic Sunyaev-Zel'dovich effect. We use iCosmo\footnote{\url{http://www.icosmo.org}, \cite{Refregier:2011}}, Healpix \citep{pixel:healpix}, iSAP\footnote{\url{http:/www.cosmostat.org/isap.html}}, 2MASS\footnote{\url{http://www.ipac.caltech.edu/2mass/}}, \emph{WMAP} \footnote{\url{http://map.gsfc.nasa.gov}} and NVSS data\footnote{\url{http://heasarc.gsfc.nasa.gov/W3Browse/all/nvss.html}}
 and the Galaxy extinction maps of \cite{Schlegel:1997yv}.
 This research is in part supported by the European Research Council grant SparseAstro (ERC-228261) and by the Swiss National Science Foundation
(SNSF).

\appendix 

\section{Simulations using the Planck Sky Model  }
\label{sec:simupsm}

The Planck Sky Model (PSM)\footnote{For more details please visit the PSM website: {\it http://www.apc.univ-paris7.fr/~delabrou/PSM/psm.html}.} \citep{PSM12} models the instrumental noise, 
the beams and the astrophysical foregrounds in the frequency range that is probed by WMAP and Planck. The simulations were obtained as follows:
\begin{itemize}
\item{\it Frequency channels:} the simulated data contains the $14$ WMAP and Planck frequency channels ranging from $23$ to $857$GHz. The frequency-dependent beams are assumed to be isotropic Gaussian PSFs.
\item{\it Instrumental noise:} instrumental noise has been generated according to a Gaussian distribution, with a covariance matrix provided by the WMAP (9-year) and the Planck consortia.
\item{\it Cosmic microwave background:} the CMB map is drawn from a Gaussian random field with WMAP 9-year best-fit theoretical power spectrum (from the $6$ cosmological parameters model). No non-Gaussianities, such as lensing or ISW effects, have been added to the CMB map. 
\item{\it Synchrotron:} this emission arises from the acceleration of the cosmic-ray electrons in the magnetic field of our Galaxy. It follows a power law with a spectral index that varies across pixels from $-3.4$ and $-2.3$ \citep{gauss:bennett03}. In the Planck data, this component mainly appears at lower frequency observations ({\it typ.} $\nu < 70$GHz).
\item{\it Free-Free:} the free-free emission is due to the electron-ion scattering and follows a power law distribution with an almost constant spectral index across the sky ($\sim-2.15$) \citep{Dick03}. 
\item{\it Dust emission:} this component arises from the thermal radiation of the dust grains in the Milky Way. This emission follows a gray body spectrum which depends on two parameters: the dust temperature and the spectral index \citep{Fink99}. Recent studies, involving the joint analysis of IRIS and Planck $545$ and $857$Ghz observations, show significant variations in both the dust temperature and the spectral index across the sky both on large and small scales \citep{PER_Dust}.
\item{\it AME:} the AME (anomalous microwave emission) -- or spinning dust -- may develop from the emission of spinning dust grains on nanoscales. This component has a spatial correlation with the thermal dust emission but has an emissivity that roughly follows a power law in the frequency range of Planck and WMAP \citep{PER_SDust}.
\item{\it CIB:} cosmological infra-red background originates from the emission of unresolved galaxies at high redshifts.
\item{\it CO:} CO emission has been simulated using the DAME H1 line survey \citep{Dame01}.
\item{\it SZ:} the Sunyaev-Zel'Dovich effect results from the interactions of the high energy electrons and the CMB photons through inverse Compton scattering \citep{SZoriginal}. The SZ electromagnetic spectrum is well known to be constant across the sky. 
\item{\it Point sources:} these components belong to two categories of radio and infra-red point sources, which can be of Galactic or extra-Galactic origins. Most of the brightest compact sources are found in the ERCSC catalogue provided by the Planck mission \citep{PER_ERCSC}. These point sources have individual electromagnetic spectra.
\end{itemize}


\section{Details of large-scale anomalies considered in this paper}\label{app:anomalies}
We consider 5 large-scale anomaly statistics in this paper. For the details of the statistics used, we refer the reader to references in \cite{Rassat:2012} for tests relating to the low quadrupole, the quadrupole/octopole alignment and the planar octopole. For tests relating to the Axis of Evil and mirror parity, we refer the reader to reference in \cite{Rassat:axes}.  Results are presented in Table \ref{tab:quad} (low quadrupole), \ref{tab:quadoct} (quadrupole/octopole alignment), \ref{tab:planar} (planar octopole), \ref{tab:aoe} (Axis of Evil), \ref{tab:parity} (mirror parity).  For the low quadrupole we calculate  the probability for a $\chi^2$ random variable with 5 degrees of freedom to take a value less than or equal to the Planck PR1 \citep{PlanckStat} expected theoretical value $C^{\rm TH}_{\rm W9,\ell=2}= 1150.56~\mu K^2$.

\begin{table}[htbp]
   \centering
   \begin{tabular}{@{} lccc @{}} 
   \hline
Map      &Quad  &Probability & Expected\\
             &power         &                 &   Theoretical \\
             &($\mu K^2$)&($\%$)&Value ($\mu K^2$)\\
\hline
1) \\
Nilc          &       230.1  &       3.7  & \\
Sevem       &       145.5  &       1.4 & \\
Smica     &      235.3 &       3.9 & 1150.5\\
Commander &      180.8   &       2.2  &  \\
GMCA-PR1  &     283.7  &       5.8 & \\
GMCA-WPR1 	 &   188.6  &     2.4  &  \\
\hline
2) QD  subtracted \\ 
Nilc          &       214.9  &       3.2 &  \\
Sevem       &       140.1  &       1.2 &  \\
Smica     &       219.9    &       3.4 &  1150.5\\
Commander &       182.3 &      2.2 &  \\
GMCA-PR1  &       254.0 &       4.6  & \\
GMCA-WPR1 	 &       169.8 &       1.9 &   \\
\hline
3) QD and ISW subtracted \\(measured amplitude\\ 2MASS + NVSS)\\
Nilc           &       167.1   &      3.0 &  \\
Sevem        &       114.7    &       1.3  & \\
Smica        &       179.9   &       3.5  & 932.2\\
Commander &       140.0 &       2.0 &  \\
GMCA-PR1  &   215.0 &       5.1  & \\
GMCA-WPR1 &       143.1 &   2.1 &     \\

\hline 
4) kDq, ISW and kSZ\\
subracted\\
Nilc           &       165.7   &      2.9 &  \\
Sevem        &       113.8    &       1.3  & \\
Smica        &       178.4   &       3.4  & 932.2\\
Commander &       139.0 &       2.0 &  \\
GMCA-PR1  &   212.7 &       5.0  & \\
GMCA-WPR1 &       141.7 &   2.1 &     \\   \end{tabular}
   \caption {Quadrupole ($\ell=2$) power and corresponding probability of the quadrupole power being so low. 1):  For 6 different CMB maps. 2): After subtraction of the kinetic Doppler quadrupole (kDq).
   3): After subtraction of the kDq and the ISW signal due to 2MASS and NVSS galaxies.  4): After subtraction of the kDq, ISW and kSZ signals. Probabilities are calculated using the expected theoretical value given from the Planck best fit results.} 
   \label{tab:quad}
\end{table}



 \begin{table}[htbp]
   \centering

   \begin{tabular}{@{} lcccccc @{}} 
   \hline
   Map&$\hat{n}_2\cdot \hat{n}_3$&Separation ($^\circ$)&Prob($\%$)\\
 \hline
 1) \\
Nilc          &       0.9853  &       9.8 & 1.5  \\
Sevem       &       0.9786  &       11.9   & 2.1 \\
Smica     &      0.9833  &       10.5  &  1.7 \\
Commander &      0.8694   &       29.6   &  13.1 \\
GMCA-PR1  &    0.1697  &       80.2 &  83.0\\
GMCA-WPR1 	 &    0.8857  &    27.7 &  11.4 \\
\hline
2) QD  subtracted \\ 
Nilc          &       0.9997  &       1.4  &  0.02 \\
Sevem       &       0.9715  &       13.7 &  2.8 \\
Smica     &       0.9973  &       4.2 &  0.27   \\
Commander &        0.8748 &       29.0  & 12.5 \\
GMCA-PR1  &      0.9826 &       10.7 & 1.7 \\
GMCA-WPR1 	 &    0.9801 &     11.4 &   2.0 \\
\hline
3) QD and ISW subtracted \\(measured amplitude\\ 2MASS + NVSS)\\
Nilc           &       0.8495   &       31.8 &  15.1 \\
Sevem        &       0.8686  &        28.7  & 13.1 \\
Smica        &       0.8947   &       26.5  & 10.5 \\
Commander &       0.9117  &       24.3 & 8.8  \\
GMCA-PR1  &   0.4753 &       61.6  & 52.5 \\
GMCA-WPR1 &   0.8842 &   27.9   &  11.6\\
 \hline
 4) kDq, ISW, kSZ \\
 subtracted\\
 Nilc           &       0.8371  &       33.2 &  16.3 \\
Sevem        &       0.8620  &        30.5  & 13.8 \\
Smica        &       0.8859   &       27.7  & 11.4 \\
Commander &       0.9069  &       24.9 & 9.3  \\
GMCA-PR1  &   0.4787 &       61.4  & 52.1 \\
GMCA-WPR1 &   0.8691 &   29.6   &  13.1\\
   \end{tabular}
\caption{The scalar product of the preferred axes of the quadrupole and octopole ($\hat{n}_2\cdot \hat{n}_3$), its corresponding separation ($^\circ$) and the probability (\%) of having such a low separation. Note: the theoretically allowed range is $[0^\circ-90^\circ$] - since the axes are not vectors. See references in \cite{Rassat:2012} for details of the statistics.}
   \label{tab:quadoct}
\end{table}


\begin{table}[htbp]
   \centering
   \begin{tabular}{@{} lcc @{}} 
   \hline
   Map & `t' value &Probability (\%)\\
   \hline
1) \\
Nilc          &       0.9184  &       15.6 \\
Sevem       &       0.8775  &         27.3 \\
Smica     &      0.9189    &          15.5 \\
Commander &     0.8728  &     28.6 \\
GMCA-PR1    &  0.9067  &        19.4 \\
GMCA-WPR1 & 0.9444  &       9.3 \\
\hline
2) QD  subtracted \\ 
Nilc          &     -&-\\
Sevem       &  -&- \\
Smica         &  -&-\\
Commander &  -&-\\
GMCA-PR1    & -&-\\
GMCA-WPR1 &  -&-\\
\hline
3) QD and ISW subtracted \\(measured amplitude\\ 2MASS + NVSS)\\
Nilc          &       0.9173  &       16.3\\
Sevem       &        0.8482  &         35.2 \\
Smica     &      0.9298   &          13.1 \\
Commander &    0.8458  &          35.8\\
GMCA-PR1    &  0.8458  &         14.1 \\
GMCA-WPR1 &  0.9540 &       7.4 \\

 \hline
 4) kDq, ISW and kSZ\\
 subtracted\\
 Nilc          &       0.9148  &       17.0\\
Sevem       &        0.8443  &         36.0 \\
Smica     &      0.9277   &          13.1 \\
Commander &    0.8416  &          36.8\\
GMCA-PR1    &  0.9230  &         14.7 \\
GMCA-WPR1 &  0.9524 &       7.6 \\
   \end{tabular}
   \caption{The `t' value for the octopole as defined in \cite{OctPlanarity} using \emph{nside=128}, calculated from the observed CMB maps (1) and after subtraction of the kDq and ISW field due to 2MASS and NVSS galaxies (3) and after subtracton of the kDq, ISW and kSZ fields (4). The probability is determined from 5000 Monte-Carlo simulations. Subtraction of only the kDq (2) does not affect the octopole planarity.}
   \label{tab:planar}
\end{table}


\begin{table}[htbp]
   \centering
   \begin{tabular}{@{} lccc @{}} 
   \hline
   Map & Angle & Prob (\%)\\ 
  
    \hline
1) \\
Nilc          &       52.1  &       22.9 \\
Sevem       &       53.6 &       28.3\\
Smica     &      51.5    &       21.1 \\
Commander &     58.0  &       49.4 \\
GMCA-PR1    &  51.4  &       20.7 \\
GMCA-WPR1 & 51.3  &   20.5 \\
\hline
2) QD  subtracted \\ 
Nilc          &       52.5  &       24.2 \\
Sevem       &       57.8 &       48.1 \\
Smica     &      21.0    &       0.06 \\
Commander &     67.0  &       88.6  \\
GMCA-PR1    &  51.6  &       21.4 \\
GMCA-WPR1 & 51.6  &   22.2 \\
\hline
3) QD and ISW subtracted \\(measured amplitude\\ 2MASS + NVSS)\\
Nilc          &       64.0  &       77.9  \\
Sevem       &       61.5  &       65.6\\
Smica     &      63.6 &       75.9 \\
Commander &     22.7  &       0.16 \\
GMCA-PR1    &  64.2  &    78.3 \\
GMCA-WPR1 & 70.0 &   96.2 \\
 \hline

 4) kDq, ISW and kSZ\\
 subtracted\\
 Nilc          &       62.2  &       69.1  \\
Sevem       &       59.6  &       56.4\\
Smica     &      63.4 &       75.0 \\
Commander &     50.4  &       17.9 \\
GMCA-PR1    &  63.9  &    77.3 \\
GMCA-WPR1 & 69.8 &   95.6 \\   \end{tabular}
   \caption{AoE mean interangle (between modes $\ell=2-5$) and its corresponding probability determined from 5000 Monte-Carlo simulations.}
   \label{tab:aoe}
\end{table}


\label{sect:parity}
 
\begin{table}[htbp]
   \centering{
   \begin{tabular}{@{} lccc @{}} 
Map&$S{_-}$&$S{_+}$\\
\hline 1) \\
Nilc                    & 3.11 (19.7\%)&2.98 (12.1\%)\\
Sevem                &  3.28 (10.6\%)&2.95 (13.8\%)\\
Smica                 & 2.80 (47.4\%)& 3.04(8.9\%)\\
Commander        &  3.39 (7.1\%)&2.99  (11.3\%)\\
GMCA-PR1       & 2.75 (53.8\%)&3.05 ( 8.3\%)\\
GMCA-WPR1   & 2.83 (43.9\%)&3.04 (9.0\%)\\
\hline
2) QD  subtracted \\ 
Nilc          &       3.19   (14.6\%)& 2.98  (12.1 \%)\\
Sevem       &        3.40  (7.0\%)& 3.12  (5.6\%)\\
Smica     &       2.88     (39.9 \%)& 3.05  (8.7\%)\\
Commander &     3.56  (3.4\%)& 3.08  (7.0\%)\\
GMCA-PR1    &  2.78  (50.9 \%)& 3.05 (8.3\%)\\
GMCA-WPR1 & 2.91   (37.1\%)& 3.04 (9.0\%)\\
\hline
3) QD and ISW subtracted \\(measured amplitude\\ 2MASS + NVSS)\\
Nilc          &       3.38  (7.3\%)&2.90 (17.7\%)\\
Sevem       &        3.46  (5.6\%)&2.54  (56.9\%)\\
Smica     &       3.22   (13.2\%)& 3.05(9.12\%)\\
Commander &     3.57  (3.3\%)&2.51  (61.1\%)\\
GMCA-PR1    &  2.92  (35.3\%)&3.08  (7.8\%)\\
GMCA-WPR1 &  3.25  (12.0\%)& 3.03 (10.0\%)\\
\hline
4) kDq, ISW and kSZ\\
subtracted\\
Nilc&3.37  (7.7\%)&2.89 (18.1\%)\\
Sevem&3.43 (6.4\%)&2.53 (58.7\%)\\
Smica&3.21 (13.7\%)&3.04 (9.3\%)\\
Commander&3.55 (3.8\%)&2.49 (63.1\%)\\
GMCA-PR1&2.91  (36.4\%)&3.07 (8.1\%)\\
GMCA-WPR1&3.24 (12.5\%)&3.02 (10.2\%)\\
\end{tabular}
}
\caption{Values of odd ($S_-$) and even ($S_+$) parity scores for $2<\ell<5$. 
   The occurrence for 5000 full-sky Gaussian random simulations is given in brackets.} 
      \label{tab:parity}
\end{table}



  \end{document}